\documentclass[pra,amsmath,amssymb, notitlepage, twocolumn,superscriptaddress,longbibliography,nofootinbib]{revtex4-1}
\usepackage[pdftex,colorlinks=true,linkcolor=darkblue,citecolor=blue,urlcolor=darkred]{hyperref}
\usepackage{braket}
\usepackage{amsmath,amsfonts, amssymb, amsthm, dsfont}
\usepackage{yfonts}
\usepackage{bm,breqn}
\usepackage{mathrsfs}
\usepackage{graphicx}
\usepackage[caption=false]{subfig}
\usepackage{verbatim}
\usepackage[export]{adjustbox}
\usepackage{tikz}
\usetikzlibrary{calc}
\usepackage{multirow}
\usepackage{color}
\usepackage[capitalise]{cleveref}

\theoremstyle{definition}

\definecolor{darkblue}{rgb}{0.,0.,0.4}
\definecolor{darkred}{rgb}{0.5,0.,0.}

\graphicspath{{figs/}{}}

\newcommand\defeq{\mathrel{\stackrel{\makebox[0pt]{\mbox{\normalfont\tiny def}}}{=}}}
\newcommand{\refeq}[1]{Eq.~(\ref{#1})}
\newcommand{\reffig}[1]{Fig.~\ref{#1}}
\newcommand{\reftab}[1]{Table~\ref{#1}}
\newcommand{\refsec}[1]{Sec.~\ref{#1}}
\newcommand{\refcite}[1]{Ref.~\cite{#1}}
\newcommand{\tr}{\text{tr}}

\newcommand{\norm}[1]{\left\lVert #1 \right\rVert}

\newcommand{\Bc}{\vcenter{\hbox{\includegraphics[height=1.5cm]{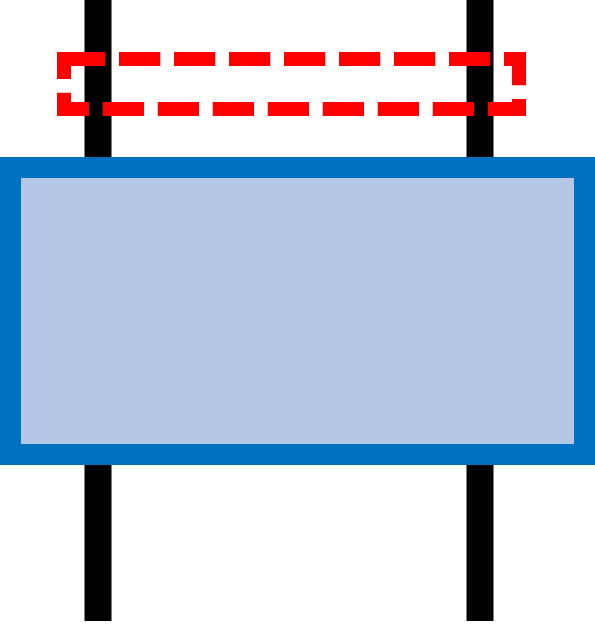}}}}
\newcommand{\Buu}{\vcenter{\hbox{\includegraphics[height=1.5cm]{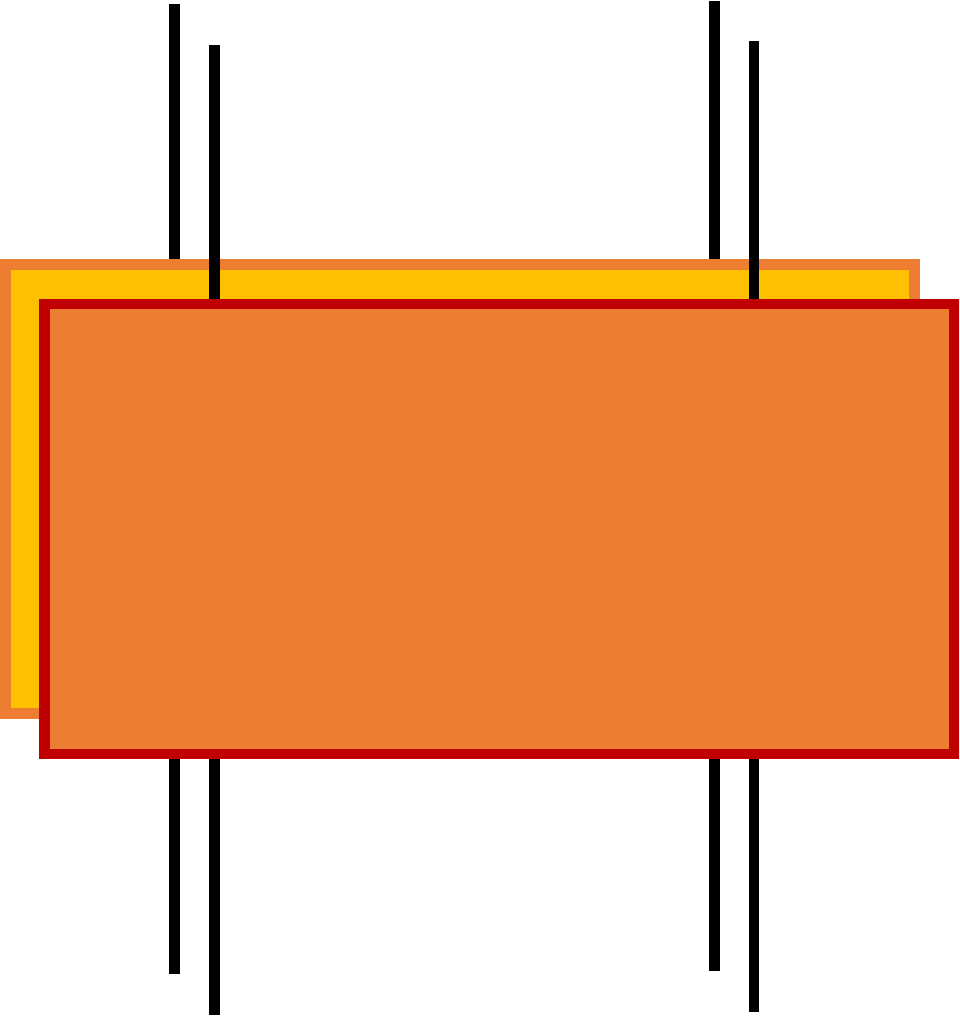}}}}
\newcommand{\Qc}{\vcenter{\hbox{\includegraphics[height=2cm]{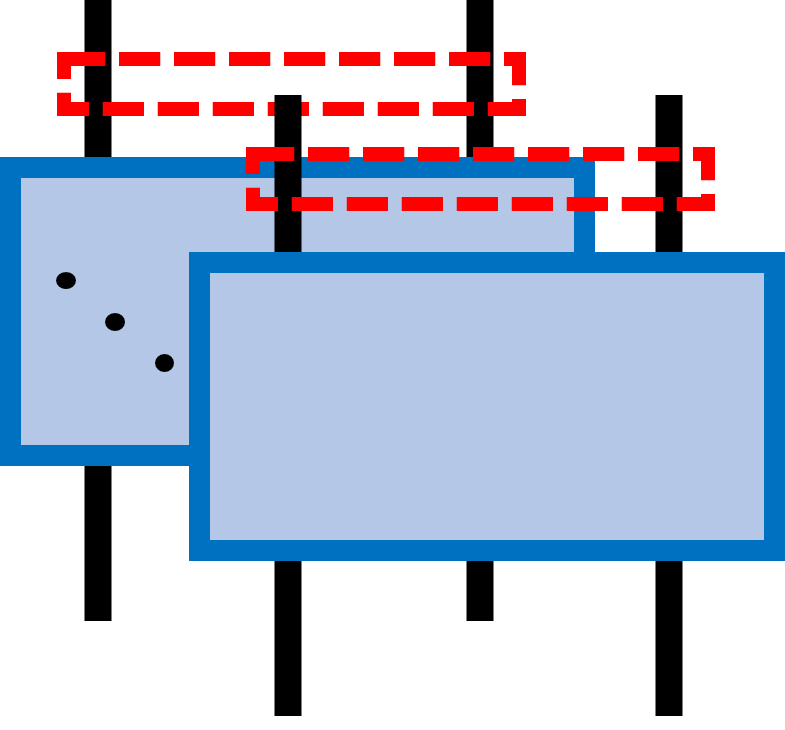}}}}
\newcommand{\Quu}{\vcenter{\hbox{\includegraphics[height=2cm]{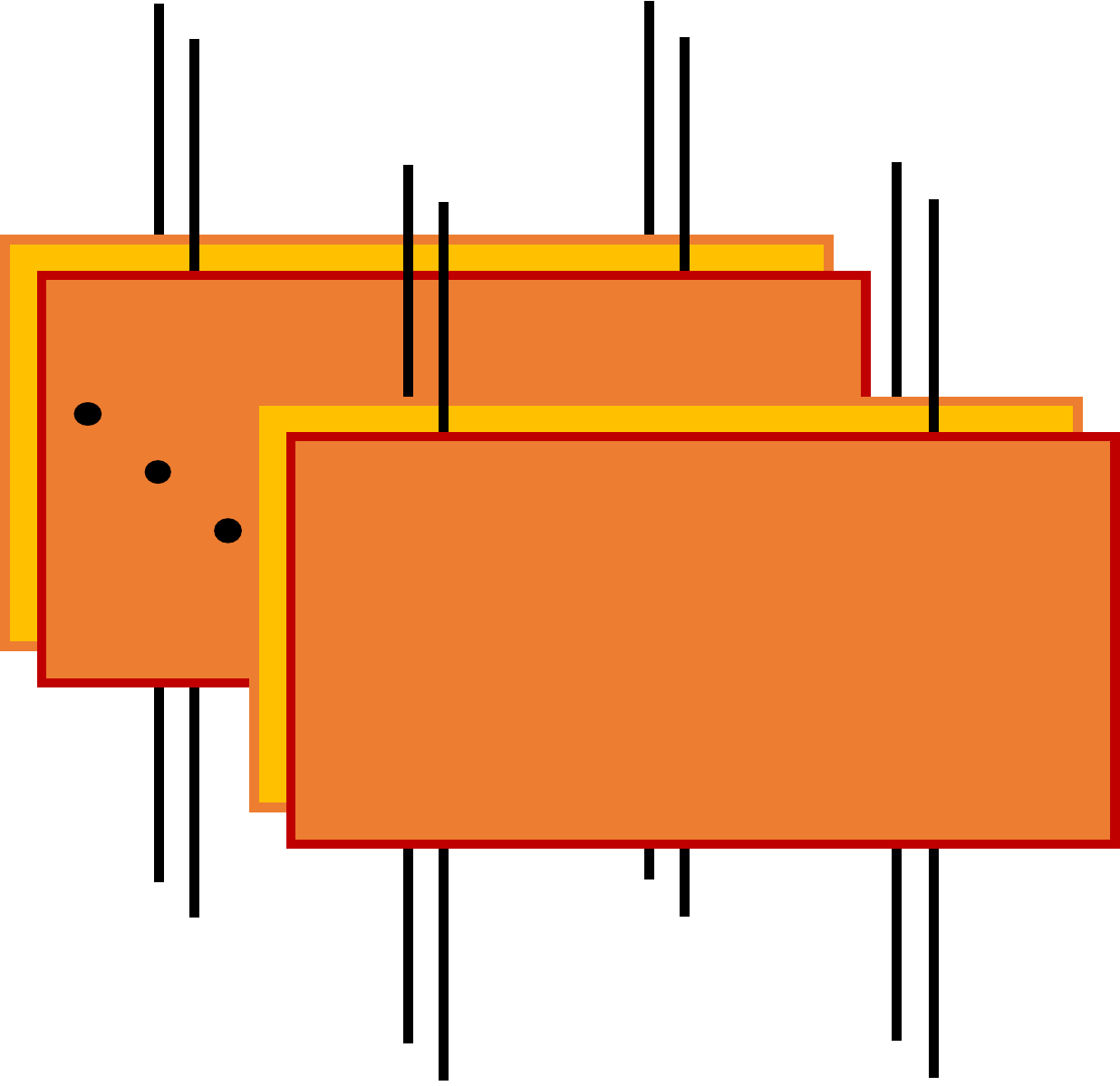}}}}
\newcommand{\Qdc}{\vcenter{\hbox{\includegraphics[height=2cm]{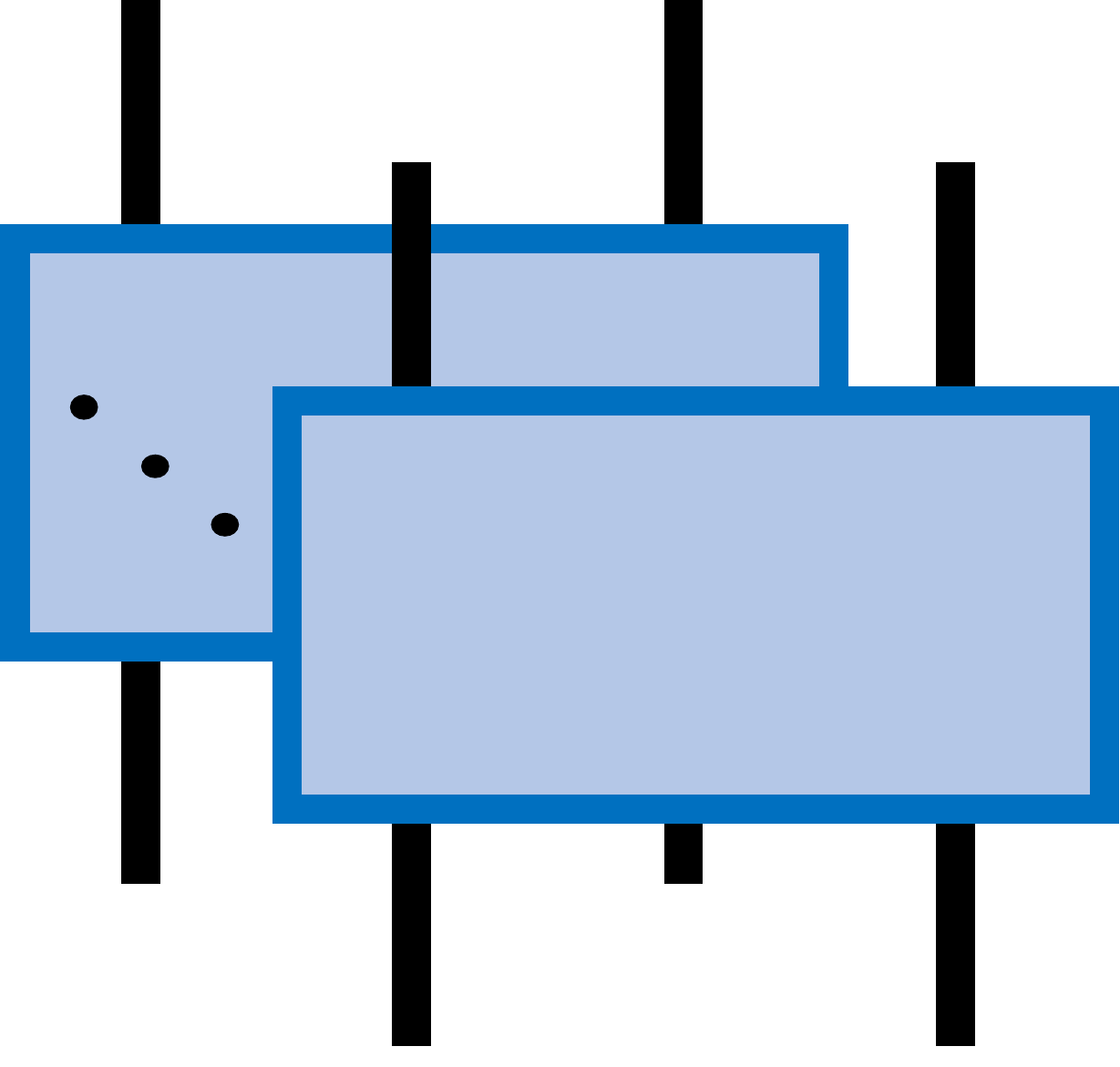}}}}
\newcommand{\Btr}{\vcenter{\hbox{\includegraphics[height=1.5cm]{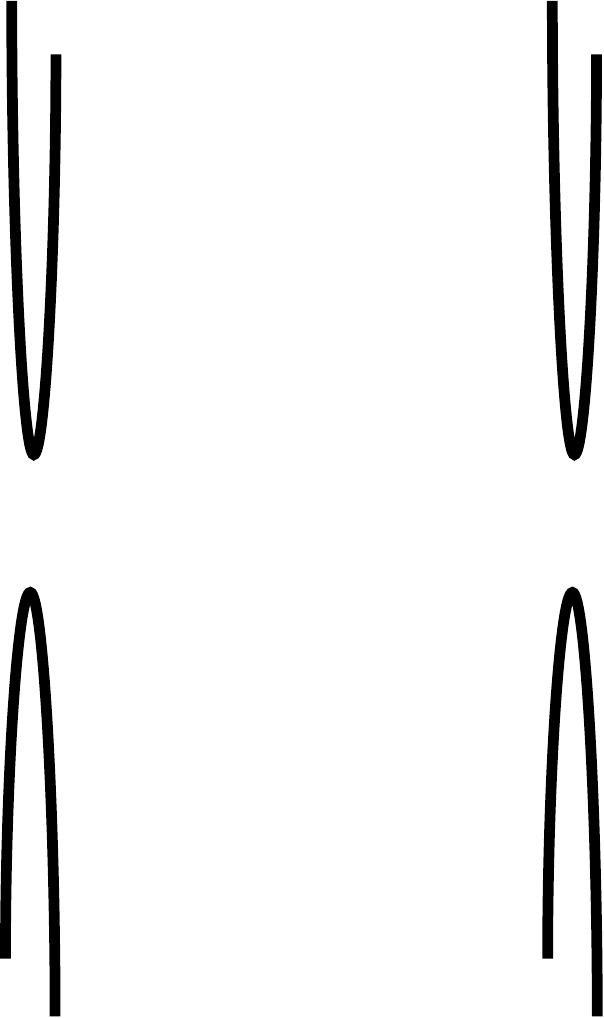}}}}
\newcommand{\Qtr}{\vcenter{\hbox{\includegraphics[height=2cm]{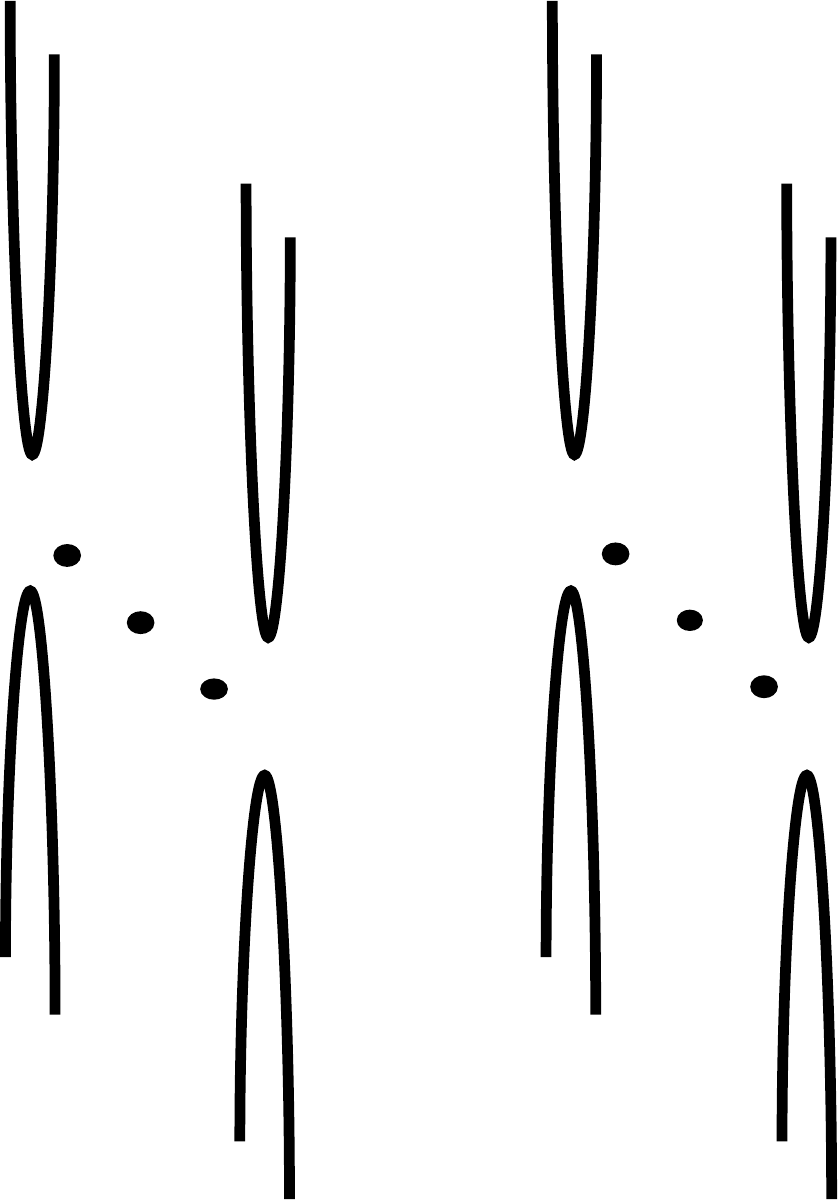}}}}
\newcommand{\Qpermid}{\vcenter{\hbox{\includegraphics[height=2cm]{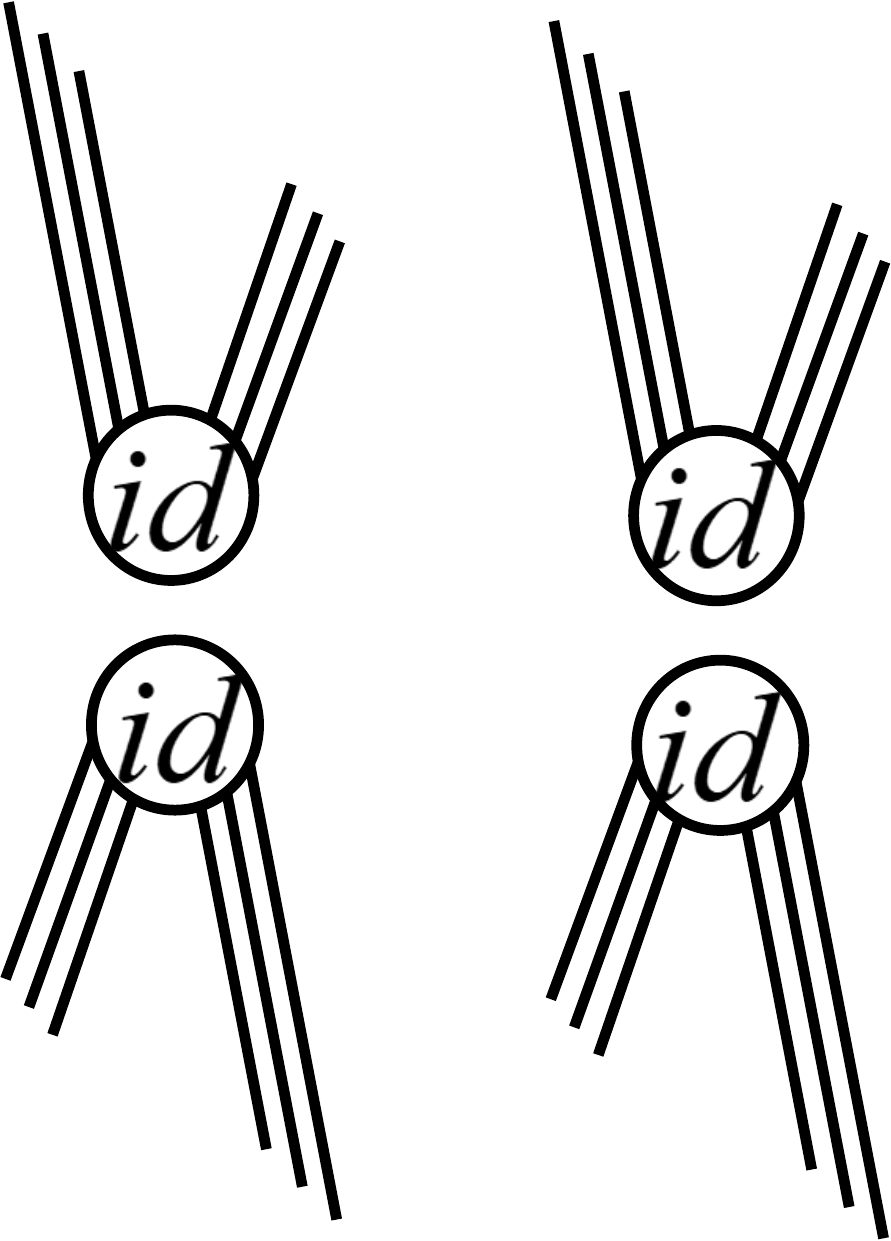}}}}
\newcommand{\Qpermg}{\vcenter{\hbox{\includegraphics[height=2cm]{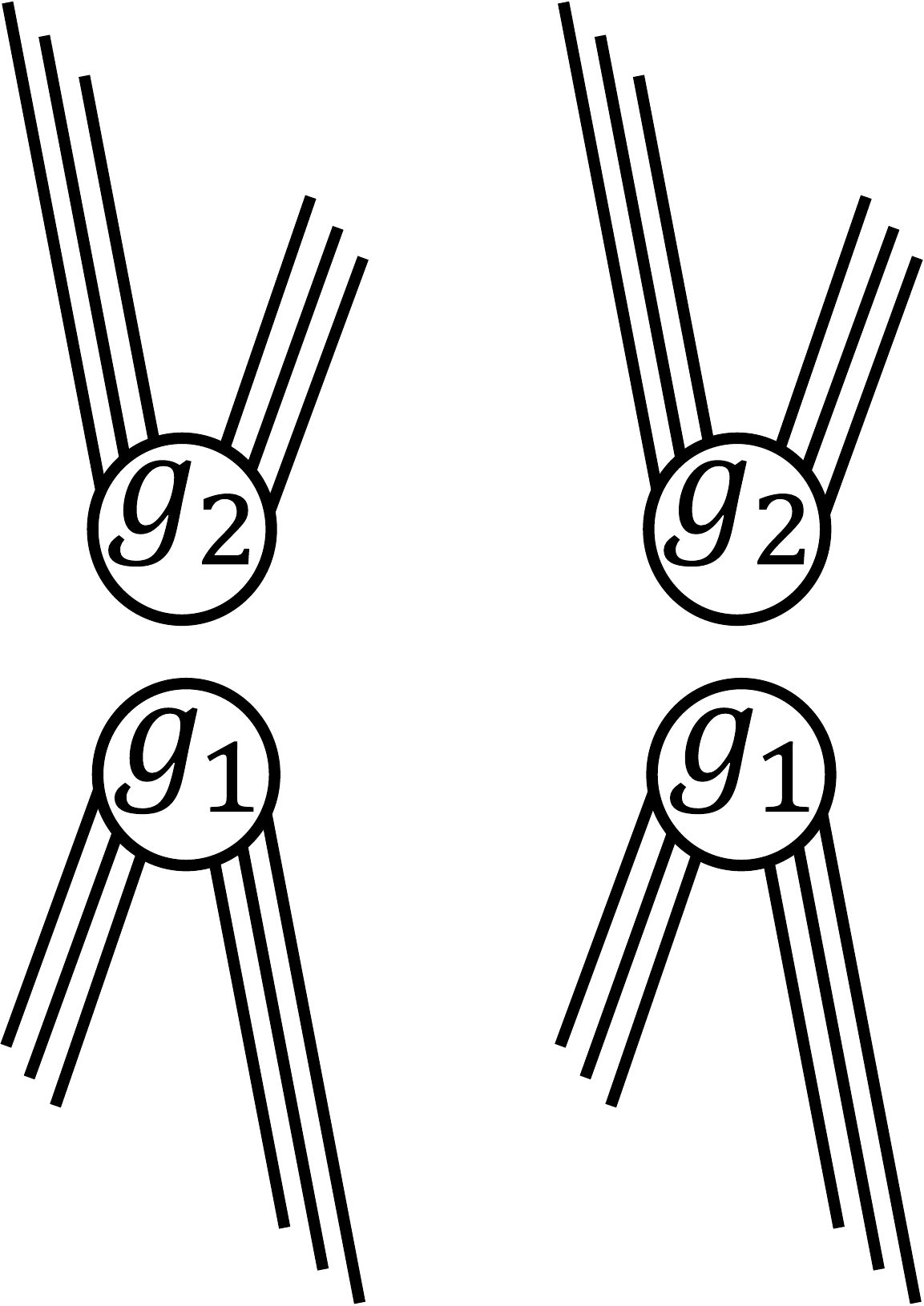}}}}
\newcommand{\Qpermh}{\vcenter{\hbox{\includegraphics[height=2cm]{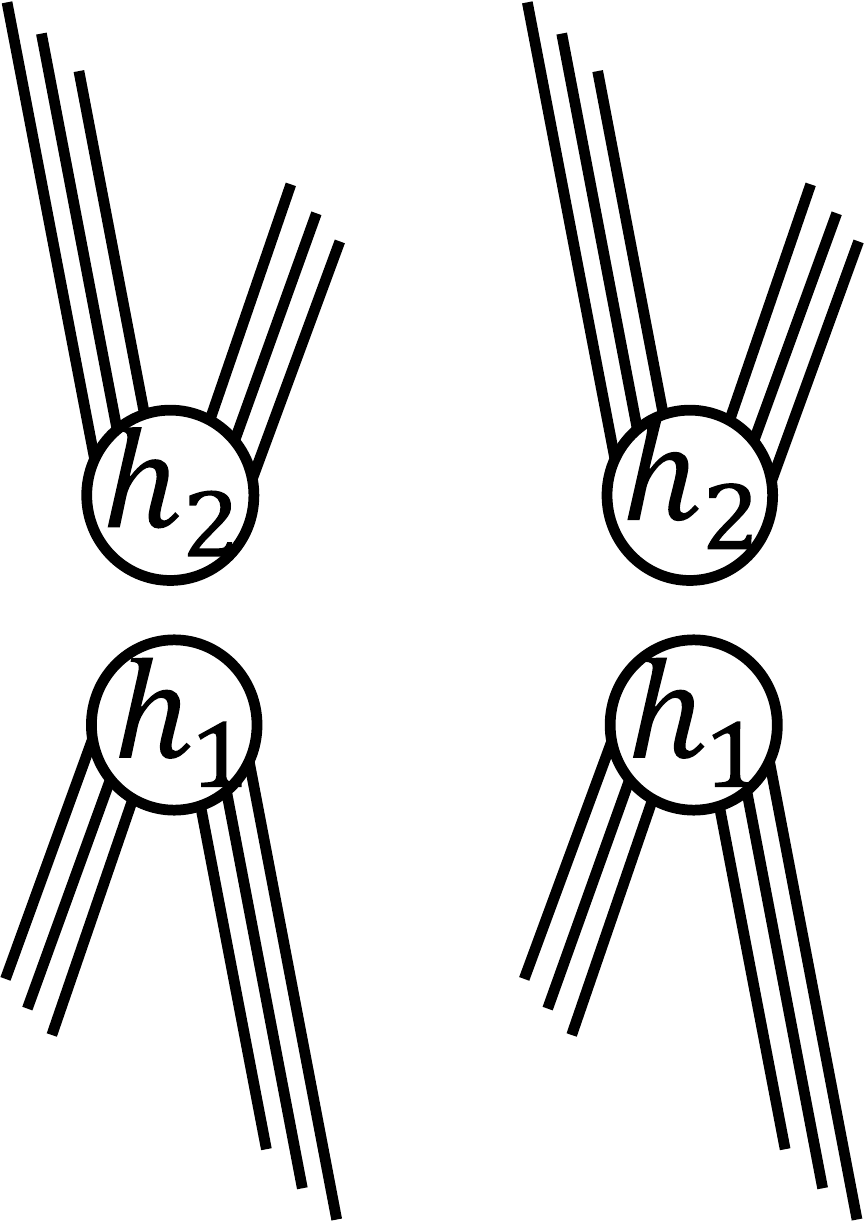}}}}
\newcommand{\Qcontraction}{\vcenter{\hbox{\includegraphics[height=1.5cm]{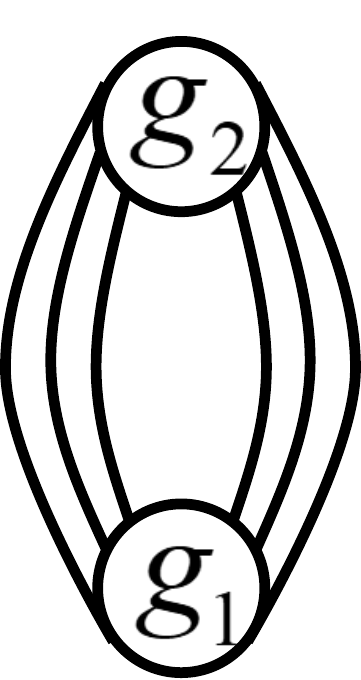}}}}
\newcommand{\Tabc}{\vcenter{\hbox{\includegraphics[width=1.3cm]{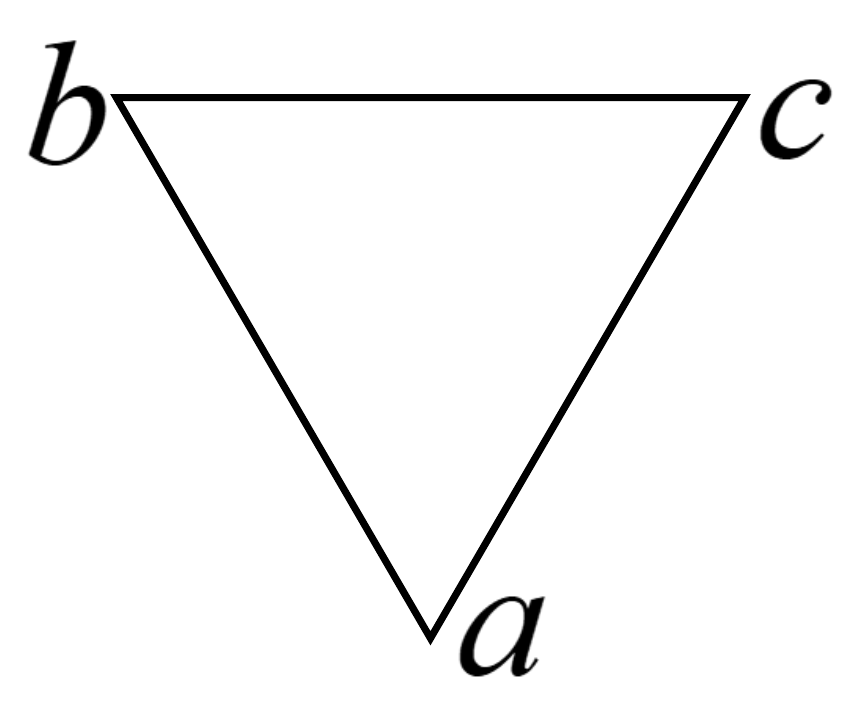}}}}
\newcommand{\Tgeneral}{\vcenter{\hbox{\includegraphics[width=2.3cm]{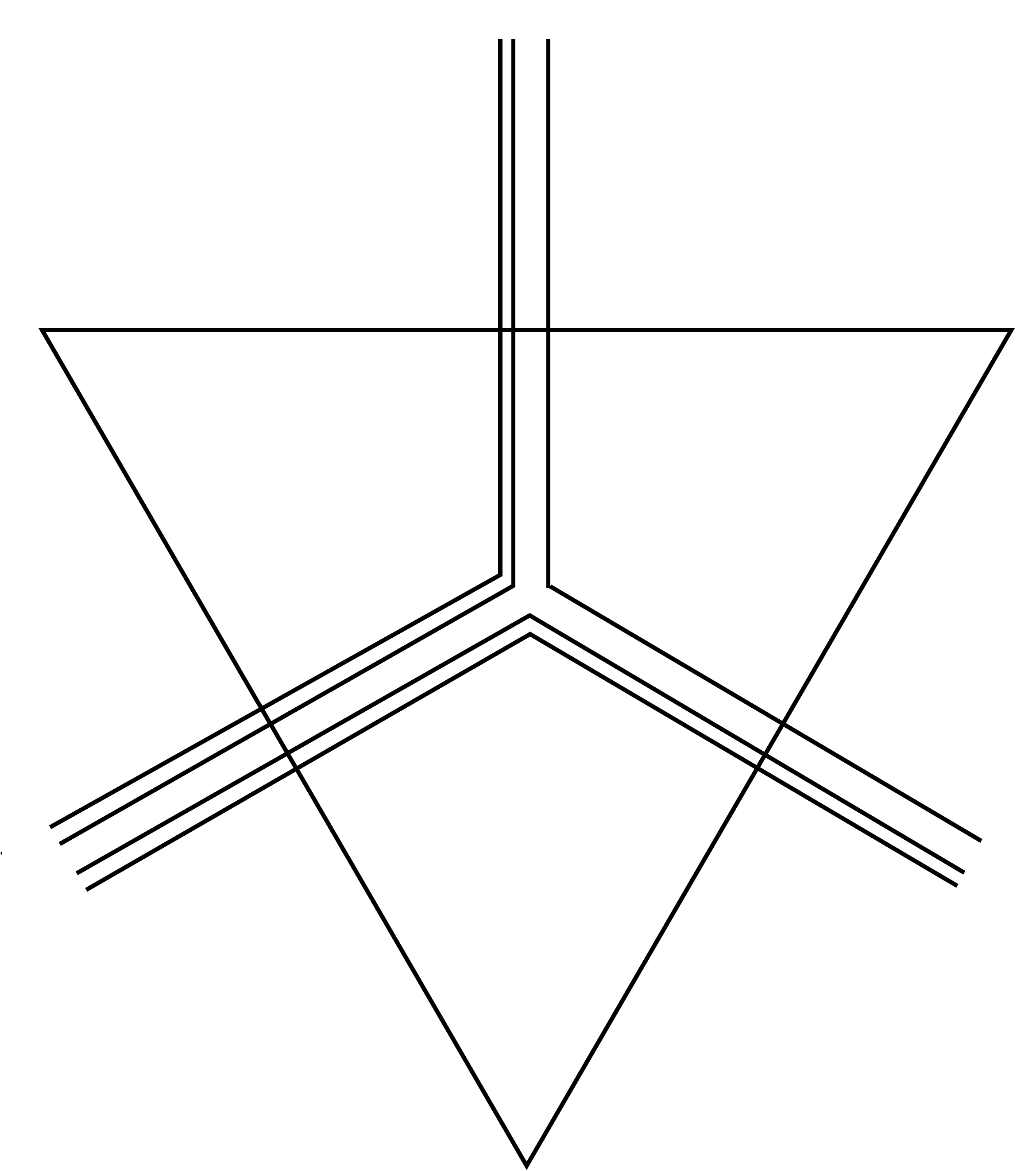}}}}
\newcommand{\Yabc}{\vcenter{\hbox{\includegraphics[width=1.3cm]{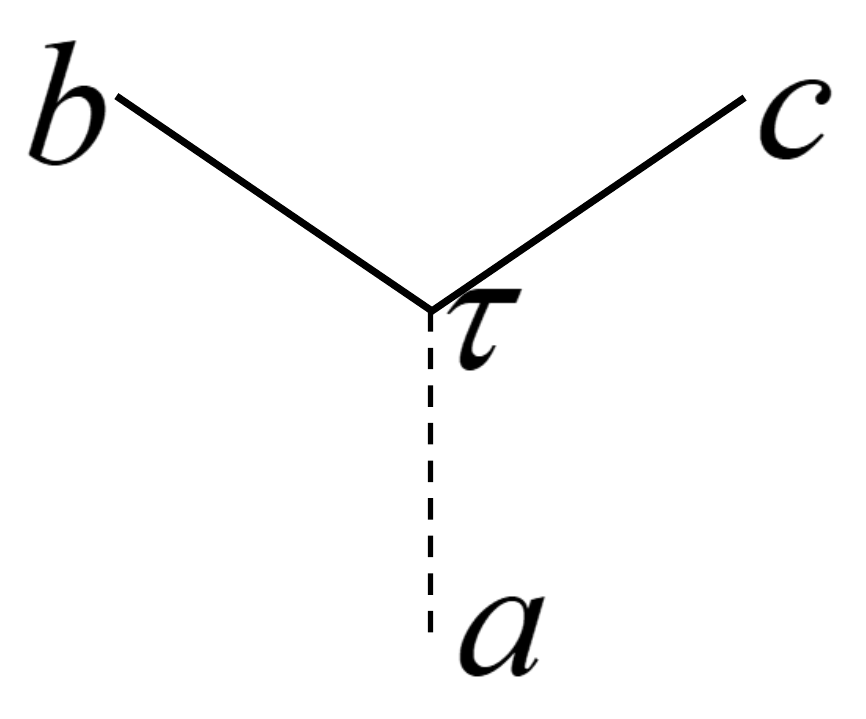}}}}
\newcommand{\Tooo}{\vcenter{\hbox{\includegraphics[width=0.9cm]{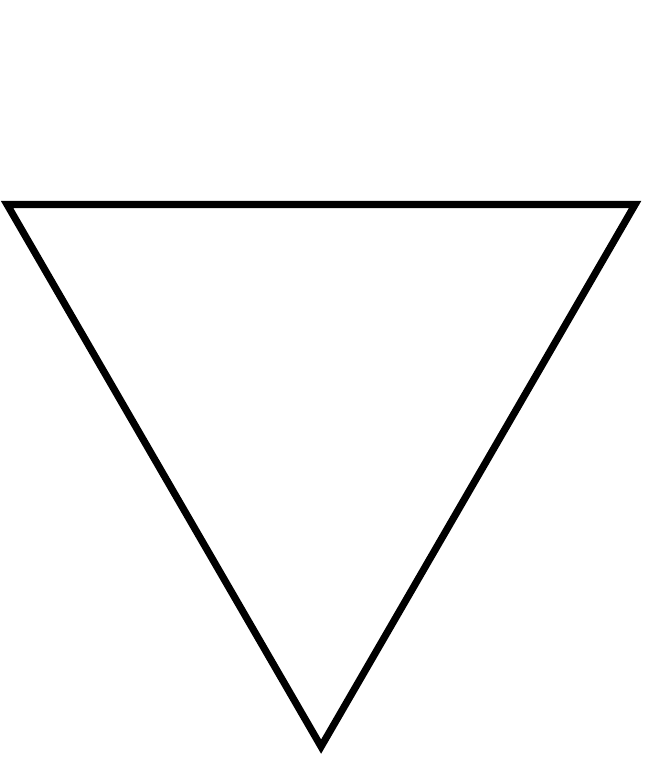}}}}
\newcommand{\Tsos}{\vcenter{\hbox{\includegraphics[width=0.9cm]{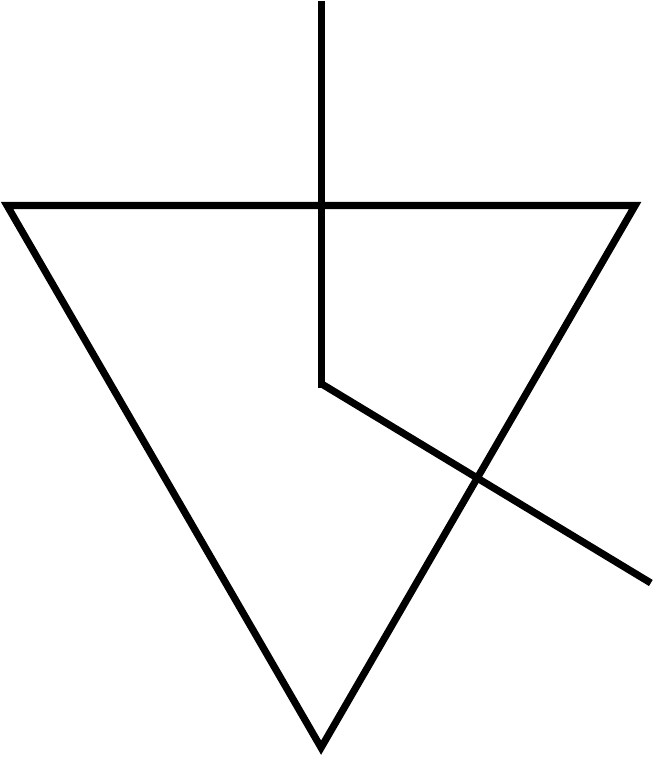}}}}
\newcommand{\Toss}{\vcenter{\hbox{\includegraphics[width=0.9cm]{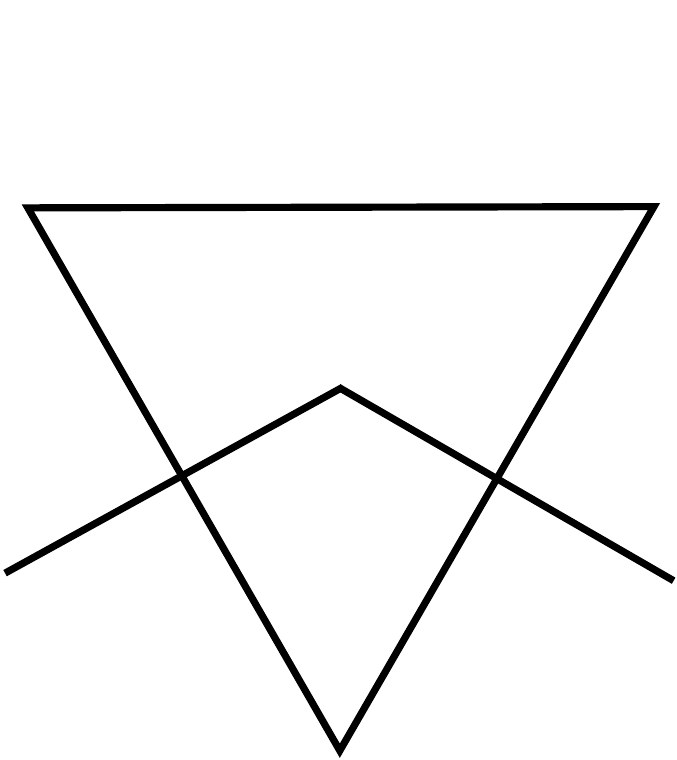}}}}
\newcommand{\Todd}{\vcenter{\hbox{\includegraphics[width=0.9cm]{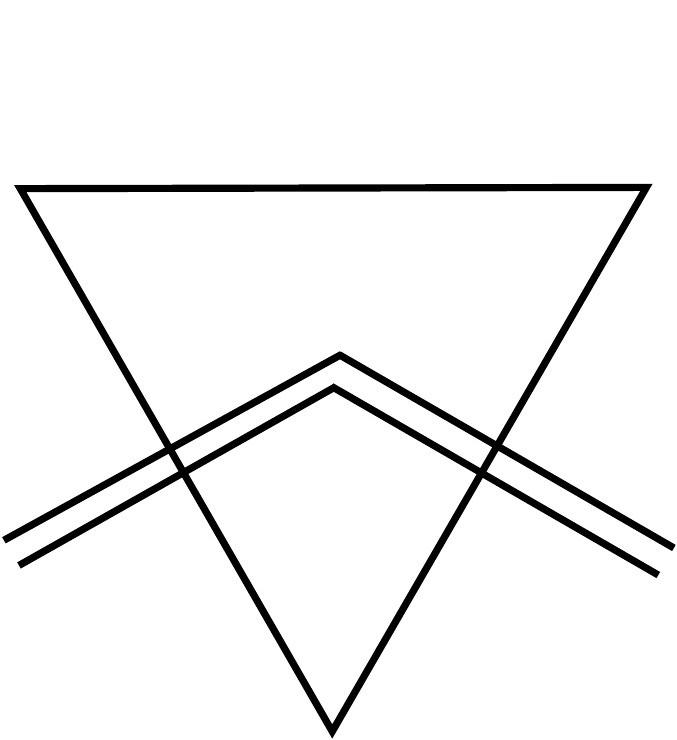}}}}
\newcommand{\Tsds}{\vcenter{\hbox{\includegraphics[width=0.9cm]{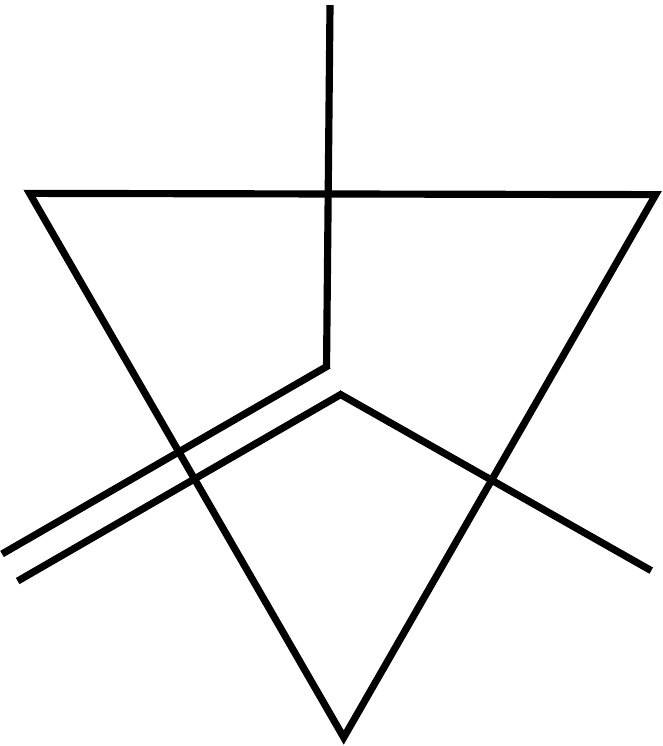}}}}
\newcommand{\Tdss}{\vcenter{\hbox{\includegraphics[width=0.9cm]{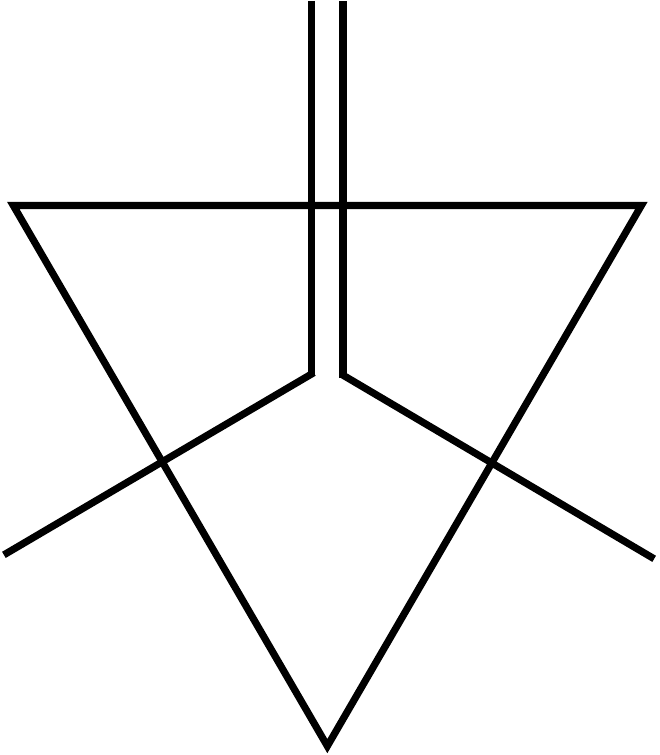}}}}
\newcommand{\Tdod}{\vcenter{\hbox{\includegraphics[width=0.9cm]{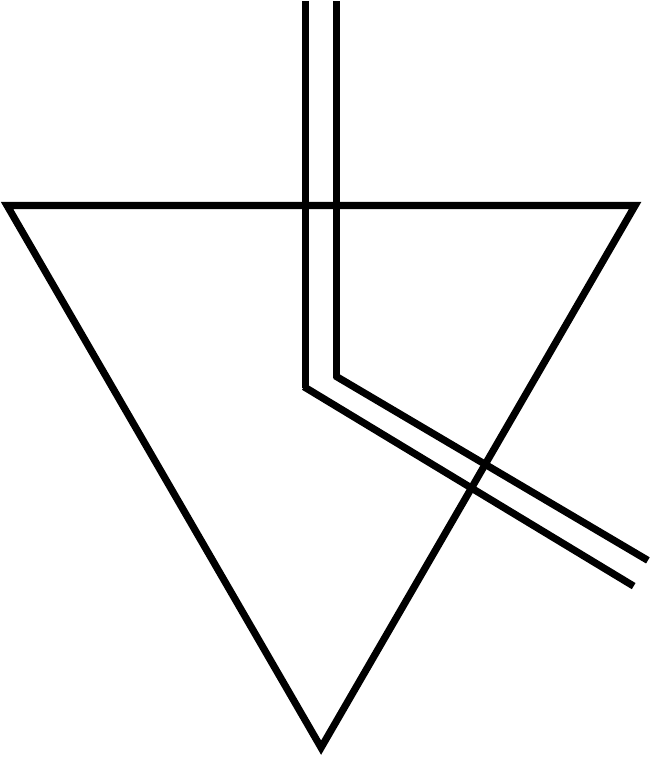}}}}

\newcommand{\sega}{\rotatebox[origin=c]{20}{$\langle$}}
\newcommand{\segb}{\rotatebox[origin=c]{-20}{$\rangle$}}
\newcommand{\segc}{\rotatebox[origin=c]{-90}{$\langle$}}

\begin{document}
\title{Entanglement Dynamics of Noisy Random Circuits}
\author{Zhi Li}\affiliation{\PI}
\author{Shengqi Sang}\affiliation{\PI}\affiliation{\UW}
\author{Timothy H. Hsieh}\affiliation{\PI}
\newcommand*{\PI}{Perimeter Institute for Theoretical Physics, Waterloo, Ontario N2L 2Y5, Canada}
\newcommand*{\UW}{Department of Physics and Astronomy, University of Waterloo, Waterloo, Ontario N2L 3G1, Canada}
\begin{abstract}
The process by which open quantum systems thermalize with an environment is both of fundamental interest and relevant to noisy quantum devices. As a minimal model of this process, we consider a qudit chain evolving under local random unitaries and local depolarization channels. After mapping to a statistical mechanics model, the depolarization (noise) acts like a symmetry-breaking field, and we argue that it causes the system to thermalize within a timescale independent of system size.  We show that various bipartite entanglement measures---mutual information, operator entanglement, and entanglement negativity---grow at a speed proportional to the size of the bipartition boundary. 
As a result, these entanglement measures obey an area law: Their maximal value during the dynamics is bounded by the boundary instead of the volume.  In contrast, if the depolarization only acts at the system boundary, then the maximum value of the entanglement measures obeys a volume law.  We complement our analysis with scalable simulations involving Clifford gates, for both one- and two-dimensional systems.

\end{abstract}
\maketitle

\section{Introduction}
A quantum system interacting with an environment generically thermalizes, and attempts toward understanding such dynamics have led to many different technical approaches \cite{davies1976quantum,lindblad1976generators,QTT1993,breuer2002theory,alicki2007quantum,rivas2012open}. Advances in quantum hardware have added further motivation to understand such dynamics, as physical systems inevitably evolve in the presence of noise and decohere without fault tolerance.  Determining if noisy devices offer a quantum advantage over classical simulation \cite{Preskill2018quantumcomputingin} benefits from an understanding of the entanglement dynamics.  If the mixed state of the system is not too entangled during the dynamics, then classical simulations may be efficient \cite{VidalMPS,MPOnumerics,PhysRevX.10.041038,PhysRevLett.93.207204,PhysRevLett.93.207205}.

Random circuits have provided a fruitful approach for studying many-body quantum dynamics of closed systems \cite{PhysRevX.7.031016}. As a toy model, generic unitary time evolution is represented by local random unitaries, admitting a mapping of the dynamics to a classical statistical mechanics (stat-mech) model \cite{ZhouNahum}.  This allows calculating many essential features of many-body quantum dynamics such as entanglement growth \cite{PhysRevX.7.031016}, spectral form factors \cite{PhysRevLett.121.060601}, out-of-time-ordered correlations \cite{PhysRevX.8.021013}, and operator growth \cite{PhysRevX.8.021014}. The random circuit can also be hybridized with measurements, yielding fascinating phenomena \cite{PhysRevX.9.031009,PhysRevB.98.205136,PhysRevB.99.224307}. 

In this work, we use the random circuit approach to study the dynamics of open quantum systems.  Such an application has already led to valuable insights in different contexts \cite{PhysRevB.102.134310,PhysRevB.103.115132,li2021robust,weinstein2022measurement}. As a minimal model for a one-dimensional (1D) quantum system inside an infinite-temperature bath, we consider a random channel circuit consisting of random local unitaries and local depolarization channels. We are interested in how the system eventually reaches equilibrium (in this case, a maximally mixed state). We map the system to a classical model of spins taking values in a permutation group. After the mapping, the effects of the environment (depolarizing channels) manifest as a \emph{permutation symmetry-breaking field} which polarizes the spins and makes the system short-range correlated, thus setting a system-size-independent timescale to reach equilibrium. More specifically, we use mutual information, operator entanglement, and entanglement negativity as diagnoses of correlations, and we study their time dependence.  These quantities show linear growth at early time, then reach their peak values and eventually drop to zero. Importantly, regardless of the depolarization strength, we argue based on the stat-mech model that the peaks are reached at a system-size-independent time, and the initial linear growth slopes are upper bounded by the size of the partition boundary. As a result, the peak values obey an area law: they are upper bounded by the size of the partition boundary as opposed to volumes of subsystems.  

This setup was considered in \refcite{MPOnumerics}, which reached the same conclusion for operator entanglement entropy based on numerics.  Here, we provide analytic arguments for this conclusion. For a complementary and scalable numerical simulation, we also consider a slightly different setup where the depolarization acts in a probabilistic fashion and the random unitaries are restricted to Clifford gates.  In this setup, we also find an area law for the entanglement peaks, in both one and two-dimensional (2D) systems.  Finally, we consider a model in which depolarization only occurs at the system boundary, and we find the entanglement peaks obey volume law, based on both analytical arguments and numerical calculations.  

\subsection{Setup}
\begin{figure}[h]
\includegraphics[width=0.6\linewidth]{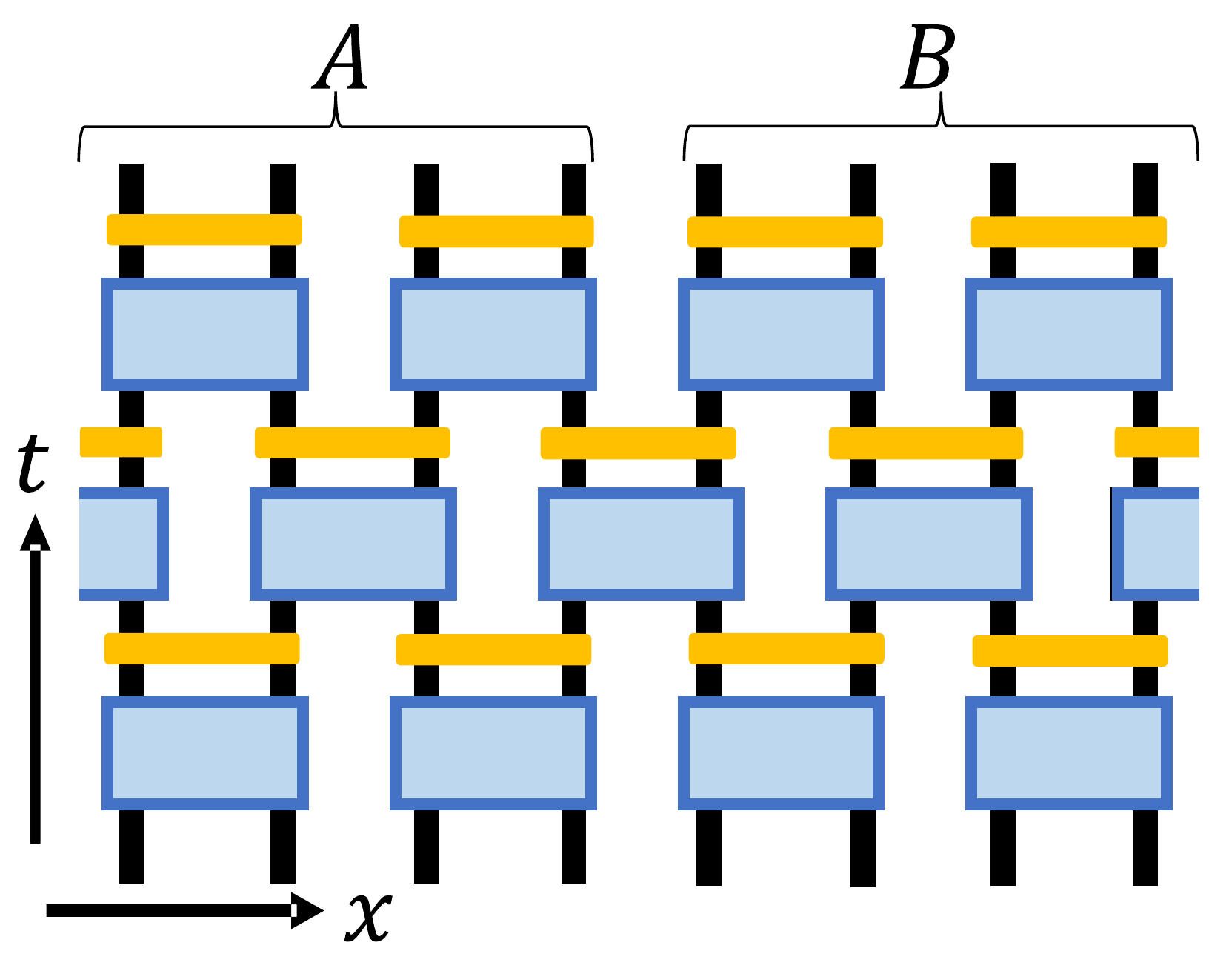}
\caption{The random circuit consists of random unitaries (blue blocks) and depolarizing channels (yellow strips).}
\label{fig-circuits}
\end{figure}

We consider a 1D qudit chain with local Hilbert space dimension $d$. The dynamics is shown in \reffig{fig-circuits}(a), where blocks are Haar random unitaries $U$ chosen independently and strips are depolarizing channels with a fixed strength parameter $p$. The combination gives the following \textit{random quantum channel}:
\begin{equation}\label{eq-channelprime}
    \Phi(\rho)=(1-p)U\rho U^\dagger+\frac{p\tr(\rho)}{d^2}\mathbb{I}_2.
\end{equation}
Here $\mathbb{I}_2$ is the maximally mixed state of two qudits. We choose the initial state to be a pure product state.

We bipartite the system into (not necessarily equal-size) subsystems $A$ and $B$ and focus on the dynamics of the Rényi-2 mutual information for concreteness:
\begin{equation}\label{eq-renyimutual}
\begin{aligned}
    I_2(A\colon B)=&S_{2,A}+S_{2,B}-S_{2,AB}\\
    =&-\log\tr\rho_A^2-\log\tr\rho_B^2+\log\tr\rho_{AB}^2,
\end{aligned}
\end{equation}
where $S_{2,*}$ is the Rényi-2 entropy of the reduced density matrices $\rho_*$ where $*$ stands for (sub)systems $A$, $B$ and $AB$.

Due to the randomness in the Haar unitaries, $I_2$ takes different values for each circuit realization and corresponding trajectory $\rho_t$. We are interested in the average over circuit realizations:
\begin{equation}\label{eq-Emutual}
    \overline{I_2}(t)=\mathbb{E}_U \left[I_2(\rho_t)\right].
\end{equation}

In the Appendix, we also consider the entanglement negativity \cite{VidalNegativity}---a mixed state entanglement measure, and operator entanglement entropy \cite{bandyopadhyay2005entangling}---the complexity of representing the density matrix as a matrix product operator. The behavior of these quantities are similar.

\section{Mapping to Statistical Mechanics Model}\label{sec-spinmodel}

The averaged mutual information \refeq{eq-Emutual} can be computed by mapping to a statistical-mechanics model, similar to Refs.~\cite{ZhouNahum,Bao,JianYou}.  In this section, we show the detailed mapping procedure and discuss some properties of the resulting stat-mech model.

\begin{figure}[h]
\includegraphics[width=\linewidth]{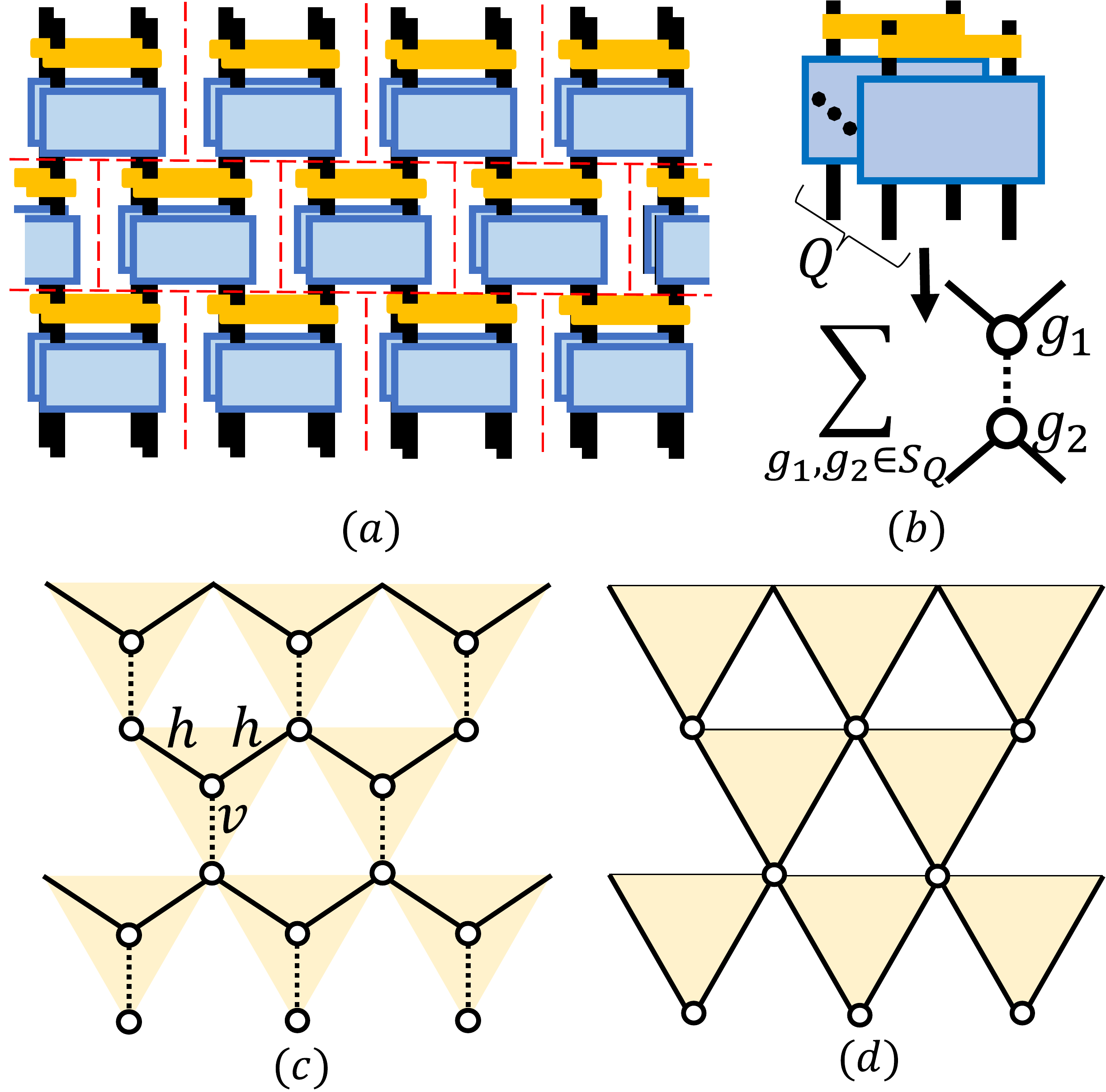}
\caption{(a) A stack of multiple copies of the circuit, decomposed (by red dashed lines) into spacetime units. (b) Haar random average via Weingarten calculus. (c) Averaged stack maps to a spin model on the honeycomb lattice. (d) Integrating out a subset of spins yields a simpler model.}
\label{fig-map}
\end{figure}

The high-level picture of the mapping is summarized in \reffig{fig-map} and are divided into the following steps:
\begin{itemize}
\item To evaluate the average of logarithms appearing in \cref{eq-renyimutual,eq-Emutual}, we use the replica trick:
\begin{equation}\label{eq-replicatrick}
    \mathbb{E}\log X=\frac{\partial}{\partial \alpha}\mathbb{E} X^\alpha\bigg\rvert_{\alpha=0}=\frac{\partial}{\partial \alpha}\log\mathbb{E} X^\alpha\bigg\rvert_{\alpha=0}.
\end{equation}
Here the last equation is due to $\mathbb{E} X^\alpha|_{\alpha=0}=1$.
    \item Each term in  the above equation involves several copies of the state, represented by stacking copies of \reffig{fig-circuits} into a multilayer circuit [\reffig{fig-map}(a)]. 
It can be decomposed into spacetime units.
Unitaries in different units are independent, while unitaries in the same unit are the same.
\item Within each unit, we average the Haar random unitary by applying a ``Weingarten calculus" transformation [\reffig{fig-map}(b)].
This transforms each unit to a different form in which two effective ``spin'' degrees of freedom taking values in $S_{Q}$, the permutation group of $Q$ elements, are summed over. Here $Q$ is the number of layers evolved; $Q=2\alpha$ for \cref{eq-renyimutual}.
\item After transforming all units, the spins reside at the lattice sites of a honeycomb lattice, giving us an effective stat-mech model on the honeycomb lattice [\reffig{fig-map}(c)]. 
\item Finally, integrate out a subset of spins and simplify the model to  \reffig{fig-map}(d). 
\end{itemize}

\subsection{Bulk theory}

Representing each $\rho$ (and $\ket{\rho}$) as a single-layer circuit (here the convention is that a pure state and its ``dual" together form a layer), traces evolved in \cref{eq-renyimutual} can be visualized as multilayer circuits. Taking the replica trick into consideration, we need to work out the traces for a generic number of layers.

To simplify the notation, we write the random channel \refeq{eq-channelprime} as:
\begin{dmath}
    \Phi(\rho)=(1-p)\Phi_U(\rho)+p\Phi_T(\rho),
\end{dmath}
where $\Phi_U$ is the random unitary channel and $\Phi_T$ is the trace channel.
Therefore,
\begin{equation}\label{eq-EQdepolar}
\begin{aligned}
  &\mathbb{E}\left(\Phi^{\otimes Q}\right)=\mathbb{E}\left[(1-p)\Phi_U+p\Phi_T\right]^{\otimes Q}\\
  =&\sum_{k=1}^Q\sum_{i_1,\cdots,i_k}(1-p)^{Q-k}p^k \mathbb{E}(\Phi_U\otimes\cdots\otimes\Phi_T\cdots\otimes\cdots).
\end{aligned}
\end{equation}
Here $\{i_1,\cdots,i_k\}$ are the positions where $\Phi_T$ appears in $\Phi_U\otimes\cdots\otimes\Phi_T\cdots\otimes\cdots$.

The only randomness here is the Haar randomness for the unitaries. We will make use of the following ``Weingarten calculus" identity for Haar average of $m$ pairs of $U$ and $U^\dagger$:
\begin{equation}\label{eq-Haaraverage}
\begin{aligned}
  \mathbb{E}\left(\Phi_U^{\otimes m}\right)=&\mathbb{E}\left(\Qdc\right)=\mathbb{E}\left(\Quu\right)\\
  =&\sum_{h_1,h_2\in S_{m}}W_m(h_1^{-1}h_2)\Qpermh.
\end{aligned}
\end{equation}
Here  the orange blocks are Haar random unitaries $U$ (and yellow for $U^\dagger$); $S_m$ is the permutation group over $m$ elements; $W_{m}()$ is the Weingarten function (it also depends on $d^2$ but we omit it). Therefore, $\mathbb{E}(\Phi_U\otimes\cdots\otimes\Phi_T\cdots\otimes\cdots)$ terms in \refeq{eq-EQdepolar} equals:
\begin{equation}\label{eq-EUdepolar}
    \sum_{h_1,h_2\in S_{Q-k}}\frac{W_{Q-k}(h_1^{-1}h_2)}{d^{2k}}\left(\Qpermh\right)\otimes\left(\Qpermid\right).
\end{equation}
In the above equation, $h_1, h_2\in S_{Q-k}$, and $id\in S_k$ (we omit the order of the tensor product to make the notation clear). 
We emphasize that the subscript of the Weingarten function are important: It indicates which group the arguments live in. If $a\in S_{m}$ and $b\in S_{n}$ are defined as acting $a$ on $\{1,2,\cdots, m\}$ and keeping $\{m+1,\cdots, n\}$ fixed (so that $a$ and $b$ are essentially the same), then it could be that $W_{m}(a)\neq W_{n}(b)$.

Plugging \refeq{eq-EUdepolar} into \refeq{eq-EQdepolar}, we see that each term in \refeq{eq-EQdepolar} will be of the form $\Qpermg$, with some $g_i\in S_{Q}$ as the combination of $h_i$ and $id$. 
However, a term with $g_1, g_2$ may come from more than one $k$ and $\{i_1,\cdots,i_k\}$. Any subset of the common fixed points of $g_1, g_2$ could come from the $\Phi_T$ part  in \refeq{eq-EQdepolar} [equivalently, the $id$ part in \refeq{eq-EUdepolar}]. So the coefficient before the $g_1, g_2$ term should be:
\begin{equation}
    \sum_{i=0}^{n_{g_1,g_2}}\binom{n_{g_1,g_2}}{i}(1-p)^{Q-i}p^i\frac{W_{Q-i}(g_1^{-1}g_2)}{d^{2i}}.
\end{equation}
Here $n_{g_1,g_2}$ is the number of common fixed points of $g_1$ and $g_2$ [for example, if $g_1=(1 2)(3)(4)$ and $g_2=(1 2 3)(4)$, then $n_{g_1,g_2}=1$]. Note that we need to slightly abuse the notation and regard $g_1^{-1}g_2$ as living in $S_{Q-i}$. This is fine since $g_1^{-1}g_2$ acts nontrivially on at most $Q-n_{g_1,g_2}$ elements and $Q-i\geq Q-n_{g_1,g_2}$. Combinatorial numbers appear because we need to pick up $i$ elements from these fixed points and assume they come from the $\Phi_T$ part and other $Q-i$ legs come from the $\Phi_U$ part. Completing the calculation, we find:
\begin{dmath}\label{eq-EEQ}
      \mathbb{E}\left(\Phi^{\otimes Q}\right)=\sum_{g_1,g_2\in S_Q}\frac{(1-p)^{Q-n_{g_1,g_2}}}{d^{2Q}}\sum_{i=0}^{n_{g_1,g_2}}\binom{n_{g_1,g_2}}{i}(1-p)^{n_{g_1,g_2}-i}p^id^{2(Q-i)}W_{Q-i}(g_1^{-1}g_2)\Qpermg
      \\\defeq\sum_{g_1,g_2\in S_Q}\frac{(1-p)^{Q-n_{g_1,g_2}}}{d^{2Q}}V_Q(g_1,g_2)\Qpermg.
\end{dmath}

Together with the cross-layer contractions
\begin{equation}\label{eq-contraction}
    \Qcontraction=d^{Q-|g_1^{-1}g_2|},
\end{equation}
[here $|g|$ is the distance between $g$ and $id$ in $S_{Q}$, which equals the minimal number of transpositions in $g$; equivalently $Q-|g|$ is the number of cycles in the cycle decomposition of $g$; for example, if $g_1=(1 2)(3)(4)$ and $g_2=(1 2 3)(4)$, then $n_{g_1,g_2}=1$, $|g_1|=1$, $|g_2|=2$], we arrive at a statistical mechanics model on the honeycomb lattice. The sum over the spins at all sites can be regarded as the \emph{partition function} of the classical stat-mech model.

On each lattice, there is a $S_{Q}$ spin $g$; the statistical weight for a spin configuration $\{g\}$ is given by:
\begin{equation}\label{eq-honey}
    \prod_v (1-p)^{Q-n_{g_1,g_2}}V_Q(g_1,g_2)\prod_h d^{-|g_1^{-1}g_2|}.
\end{equation}
Here $v$ means vertical bonds and $h$ means horizontal (zigzag) bonds. $V_Q(g_1,g_2)$ is \refeq{eq-EEQ} multiplied by $d^{2Q}$. The $1/d^{2Q}$ factor in \refeq{eq-EEQ} is exactly canceled by $d^Q$ factors in \refeq{eq-contraction} since each vertical bond corresponds to two  horizontal bonds.

Next, we integrate out spins in the middle of downward (yellow) triangles, obtaining a spin model on a (rotated) square lattice \reffig{fig-map}(d). Now the statistical weight for a spin configuration is the product of all downward triangles, where each triangle contributes:
\begin{equation}\label{eq-Tformula}
     \Tabc=\sum_{\tau\in S_{Q}}\Yabc=(1-p)^{Q-n_a}K_p(a,b,c).
\end{equation}
Here $a,b,c\in S_Q$; $n_a$ is the number of fixed points of $a$; and
\begin{equation}
   K_p(a,b,c)=\sum_{\tau\in S_{Q}} d^{-|\tau^{-1}b|-|\tau^{-1}c|}(1-p)^{n_a-n_{\tau,a}}V_Q(\tau,a).
\end{equation}

\subsection{Boundary conditions}

Recall that the purpose of drawing \reffig{fig-map}(a) is to calculate the traces in \refeq{eq-renyimutual}. Therefore, at the upper boundary of the $Q$-layer circuit, different layers need to be suitably contracted with each other according to the trace and replica structure dictated by \cref{eq-renyimutual,eq-replicatrick}. This manifests as fixed boundary conditions in the single-layer spin model. 

More precisely, for $I_2(A\colon B)$, if the replica number $\alpha=1$, then the first term amounts to fixing spins above $B$ region to $a=id_2$ and fixing spins above $A$ region to $b=(1,2)$; the second term is similar; the third term amounts to fixing spins to $b$ everywhere. If $\alpha\geq 2$, then we need to repeat the above pattern $\alpha$ times, so $a=id_Q$, $b=(1,2)(n+1,n+2)\cdots(Q-1,Q)$. 
At the end, \refeq{eq-Emutual} becomes
\begin{equation}\label{eq-FI}
 \overline{I_{2}(A\colon B)}
 =-2\frac{\partial}{\partial Q}\left(\log\mathcal{Z}_{ba}+\log\mathcal{Z}_{ab}-\log\mathcal{Z}_{bb}\right).
\end{equation}
where the subscripts  indicate different boundary conditions in the partition function $\mathcal{Z}$. 

We note that $\mathcal{Z}_{bb}$ is also related to the Rényi-2 entropy of the whole system via the following formula:
\begin{equation}\label{eq-S2whole}
    \overline{S_{2}}=-2\frac{\partial}{\partial Q}\log \mathcal{Z}_{bb}.
\end{equation}

The bottom boundary corresponds to the initial state, which we choose to be a product pure state. Under the Haar average \refeq{eq-Haaraverage}, we should attach a state $\ket{\psi}$ to the open ends at the bottom, which effectively leaves the permutations alone (basically because $\braket{\psi|\psi}=1$). Therefore, the boundaries at the bottom are always free.\footnote{On the contrary, if the initial state is the maximally mixed state, then the bottom boundary should obey a fixed boundary condition where all spins are fixed to $id$.}


\subsection{Noise as symmetry breaking}

Before analyzing the stat-mech model quantitatively, let us consider how the symmetry changes with $p$.

If $p=0$, then since both the Weingarten functions $W(~)$ and $|~|$ are invariant under conjugation, the weights have an $S_{Q}\times S_{Q}$ ``spin rotation" symmetry acting as independent left/right group multiplication. However, if $p>0$, then to keep various $n_{g_1,g_2}$ and $n_g$ invariant, only the diagonal $S_{Q}$ group survives, acting by conjugation. Hence, the depolarization (nonzero $p$) partially breaks the ``spin rotation" symmetry. Note that if $Q=2$, then the above statement should be slightly modified since $S_2$ has a nontrivial center (which is itself). The symmetry is $S_2$ (if $p=0$) and $\{id\}$ (if $p>0$), so there is still a symmetry-breaking effect. 

Another way to see the symmetry breaking is by noticing the $(1-p)^{Q-n_a}$ factor in \refeq{eq-Tformula}. This term suggests that the depolarization channel (noise) acts as a polarizing field: spins have energy $\propto-n_a$ and favor directions with larger $n_a$, breaking the above-mentioned ``spin rotation" symmetry.

\section{Large $d$ analysis}
The stat-mech model is still pretty complicated due to the complicated triangle weights [\refeq{eq-Tformula}]. Fortunately, in the large $d$ limit, the triangle weights greatly simplify, and one can obtain a very intuitive picture.

For example, consider the simplest case is the $d=\infty$ limit. In this case, the triangle weights are:
\begin{dmath}
\Tabc=\begin{cases}
(1-p)^{Q-n_a},& \text{if~}a=b=c\\
0,&\text{otherwise}
\end{cases},
\end{dmath}
enforcing all spins to be equal. 
The symmetry-breaking effect is very manifest in this limit. Indeed, if $p=0$, spins can take all possible directions due to the above-mentioned symmetry; if $p>0$, then the symmetry breaks and spins are polarized to $id$ (for which $n_a=Q$).

For large but finite $d$, spins can differ in orientation. 
We follow \refcite{ZhouNahum} and visualize each spin configuration as regions where spins take the same values, separated by boundaries (``domain walls") between the regions. For an edge $a$---$b$ ($a\neq b$), we draw $|a^{-1}b|$ domain walls across it, each representing a transposition in the decomposition of $a^{-1}b$ [e.g., if $a^{-1}b=(1 2)(3 4)$, then we draw two domain walls, one for $(1 2)$, one for $(3 4)$]. 
The statistical weight \refeq{eq-honey} still equals a product of all triangle weights, where the triangle weights are largely determined by the domain wall configuration.

\subsection{Triangle weights}\label{sec-Tweights}
In this subsection, we work out the triangle weights up to leading order in $1/d$. 

First of all \cite{collins2006integration}, 
\begin{equation}
   d^{2(Q-i)}W_{Q-i}(g)=O(\frac{\text{Moeb}(g)}{d^{2|g|}})+O(\frac{1}{d^{2|g|+4}}).
\end{equation}
Here,  $\text{Moeb}(g)$ is a coefficient that only depends on the nontrivial part of $g$ (it does not depend on $i$ which tells us which permutation group $g$ lives in). We only the actual value of $\text{Moeb}(g)$ in \refeq{eq-moeb2}. Therefore, to the leading order,
\begin{dmath}
    V_Q(g_1,g_2)=\sum_{i=0}^{n_{g_1,g_2}}\binom{n_{g_1,g_2}}{i}(1-p)^{n_{g_1,g_2}-i}p^i\frac{\text{Moeb}(g_1^{-1}g_2)}{d^{2|g_1^{-1}g_2|}}=\frac{\text{Moeb}(g_1^{-1}g_2)}{d^{2|g_1^{-1}g_2|}},
\end{dmath}
and therefore $K_p(a,b,c)$ equals:
\begin{dmath}
   \sum_{\tau\in S_{Q}} (1-p)^{n_a-n_{\tau,a}}\text{Moeb}(\tau^{-1}a)d^{-|\tau^{-1}b|-|\tau^{-1}c|-2|\tau^{-1}a|}.
\end{dmath}
According to the triangle inequality,
\begin{equation}
    |\tau^{-1}b|+|\tau^{-1}c|+2|\tau^{-1}a|\geq |a^{-1}b|+|a^{-1}c|,
\end{equation}
where equality holds if and only if two ``parallel" conditions are satisfied:
\begin{equation}\label{eq-eqcondition}
    |\tau^{-1}b|+|\tau^{-1}a|=|a^{-1}b|,~~\text{and}~|\tau^{-1}c|+|\tau^{-1}a|=|a^{-1}c|.
\end{equation}
Hence to the leading order,
\begin{equation}\label{eq-dummy}
    K_p(a,b,c)=d^{-|a^{-1}b|-|a^{-1}c|}\sideset{}{'}\sum_{\tau\in S_{Q}} (1-p)^{n_a-n_{\tau,a}}\text{Moeb}(\tau^{-1}a),
\end{equation}
where $\sum'$ means summation over all $\tau$ satisfying \refeq{eq-eqcondition}.

For us, the relevant  graphs are of the following form:
\begin{equation}
    \Tgeneral.
\end{equation}
We denote the number of $\sega,\segb,\segc$ as $x,y,z$. These $(x+y+z)$ lines are commuting  domain walls: each domain wall is a transposition and these transpositions have no common elements thus commuting to each other. An example satisfying this graph is the following:
\begin{equation}\label{eq-abceg}
  \begin{aligned}
b^{-1}a&=(1,2)\cdots(2z-1,2z)(2z+1,2z+2)\\
&\cdots(2x+2z-1,2x+2z),\\
c^{-1}a&=(1,2)\cdots(2z-1,2z)(2x+2z+1,2x+2z+2)\\
&\cdots(2x+2y+2z-1,2x+2y+2z). 
\end{aligned}  
\end{equation}

In this case, there are $2^z$ possibilities for $\tau$ [the variable integrated out in \refeq{eq-dummy}]: $\tau^{-1}a$ is the product of some involutions choosing from $(1,2)$, $(3,4)$, $\cdots$, $(2z-1,2z)$. Under such restriction,
\begin{equation}\label{eq-moeb2}
    \text{Moeb}(\tau^{-1}a)=(-1)^{|\tau^{-1}a|}.
\end{equation}
Moreover, notice that $n_a-n_{\tau,a}=n_a-n_{\tau^{-1}a,a}$ is the number of points invariant under $a$ but change under $\tau^{-1}a$, which is the number of fixed points of $a$ restricting on the nonfixed part of $\tau^{-1}a$. For example, if $\tau^{-1}a=(1,2)(3)(4)\in S_4$, then $n_a-n_{\tau,a}=n(a|_{\{1,2\}})$ which is just $\delta_{a(1),1}+\delta_{a(2),2}$.
Hence, in the example \refeq{eq-abceg},
\begin{equation}\label{eq-Tgeneral}
    K_p(a,b,c)=\frac{1}{d^{x+y+2z}}\prod_{i=1}^z\left[1-(1-p)^{n(a|_{\{2i-1,2i\}})}\right].
\end{equation}

To summarize: 
\begin{itemize}
\item There is a prefactor $(1-p)^{Q-n_a}$, which is the symmetry-breaking effect discussed above.
    \item At the leading order of $1/d$, commuting domain walls are effectively independent of each other. 
    \item Each $\sega$ or $\segb$ contributes $1/d$.
    \item If $p\neq 0$, then horizontal domain walls $\segc$ could exist, with an extra penalty $\frac{1-(1-p)^*}{d}$.
\end{itemize}  
As a sanity check, the contribution of $\segc$ vanishes if $p=0$, so horizontal domain walls are indeed forbidden in the $p=0$ unitary-only case. 
In \reftab{tab-weights}, we list some examples of the triangle weights. 
\begin{table}[h!]
    \centering
    \begin{tabular}{c||c|c|c}
        \hline\hline
          &$\Tooo$ &$\Tsos$ &$\Toss$  \\
         \hline
         $p\neq 0$& $(1-p)^{Q-n_a}$ & $\frac{(1-p)^{Q-n_a}}{d}$ & $(1-p)^{Q-n_a}\frac{1-(1-p)^{n_a-n_{a,c}}}{d^2}$  \\
         \hline
        $p=0$ &1 & $1/d$ & 0  \\
         \hline\hline
    \end{tabular}
    \caption{Some triangle weights for the spin model. $a, b, c$ is the spin on the bottom, left, right. We also list $p=0$  unitary-only case for comparison.}\label{tab-weights}
\end{table}

\subsection{Qualitative analysis}
The above triangle weights provide us the following physical picture of the stat-mech model. 
\begin{itemize}
    \item Each domain wall contributes at least a $1/d$ factor, making the spins try to parallel with each other (``ferromagnetic" interaction). 
    \item $p\neq 0$ introduces a factor $(1-p)^{Q-n_a}$, making the spins try to parallel to $id$ (``magnetic" field).
\end{itemize}
One immediately notices that the physics is very similar to those in the ferromagnetic Ising model with magnetic field. Indeed, for $Q=2$, the triangle weights in \reftab{tab-weights} can be equivalently summarized in terms of energies [up to $O(1/d)$] as:
\begin{equation}\label{eq-Ising}
    E=-\log d\sum_{\braket{ij}}(\delta_{\sigma_i\sigma_j}-1)-\log\frac{1}{1-p}\sum_i(\delta_{\sigma_i}-1),
\end{equation}
which is exactly the Ising model with a magnetic field on a (rotated) 2D square lattice. 

Now let us proceed to calculate the Renyi-2 entropy $I_2$. Using \refeq{eq-FI}, we find:
\begin{dmath}\label{eq-FS}
 \overline{I_{2}}=-2\frac{\partial}{\partial Q}(\log\mathcal{Z}_{ba}+\log\mathcal{Z}_{ab}-\log\mathcal{Z}_{bb})
 \approx -\log\mathcal{Z}_{ba}-\log\mathcal{Z}_{ab}+\log\mathcal{Z}_{bb}\Big\rvert_{Q=2}.
\end{dmath}
Here, recall that $a=id$ and $b=(12)(34)\cdots(Q-1,Q)$; $\mathcal{Z}_{g_1g_2}$ denotes the partition function with spins above $A/B$ regions fixed to $g_1/g_2$, respectively. 
The second line is due to the approximate independence discussed in \refsec{sec-Tweights}: $Q/2$ domain walls contribute independently, yielding $\mathcal{Z}(Q)\approx \mathcal{Z}(Q=2)^{Q/2}$ [see discussions near \cref{eq-decom1,eq-decom2} for more details].

\begin{figure}[h]
    \centering
    \includegraphics[width=0.45\textwidth]{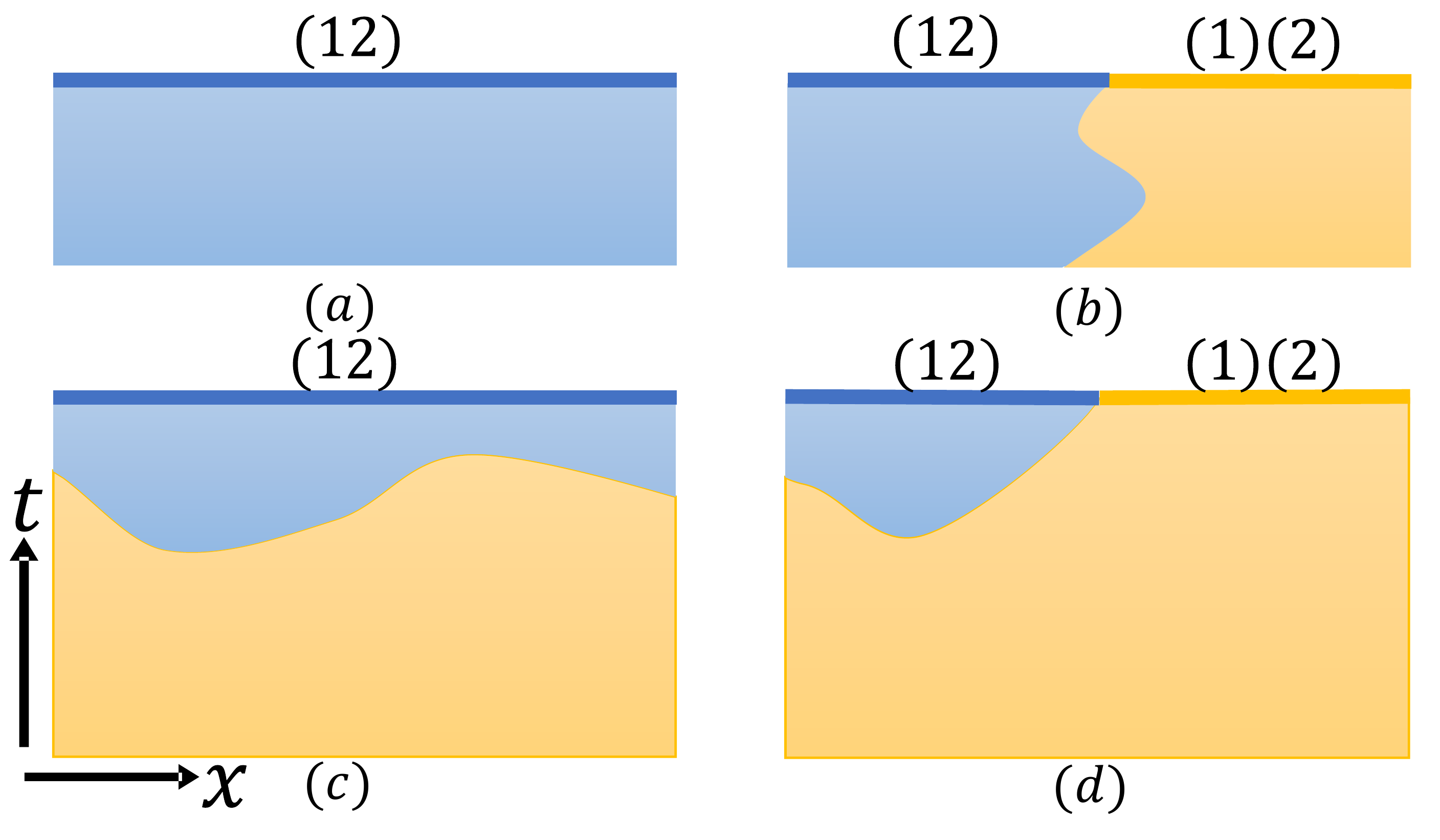}
    \caption{Domain wall configurations for  $I_2$ at (a, b) small $t$ and (c, d) large $t$. Thick lines on the upper boundary indicate fixed boundary conditions. Yellow spins are $id=(1)(2)$; blue spins are $(12)$. Domain walls in (c, d) are horizontally directed. Here $Q=2$ and there is only one domain wall.}
    \label{fig-dm}
\end{figure}

Due to the nearest-neighbor ferromagnetic interaction, spins near the upper boundary tend to be parallel to the fixed boundary conditions. However, spins deep in the bulk tend to be polarized to $id$ due to the polarization field. Therefore, which configurations dominate depends on the time $t$; see \reffig{fig-dm}.

More precisely, for small $t$, dominant configurations should have no domain walls  or vertical domain walls only (due to the energy penalty $1/d$ for horizontal domain walls); see \reffig{fig-dm}(a) for $\mathcal{Z}_{bb}$ and \reffig{fig-dm}(b) for $\mathcal{Z}_{ba}$. When $t$ is large, the above configurations are not economical anymore. Instead, we need to consider domain walls separating the upper and lower boundaries, see Figs.~\ref{fig-dm}(c) and \ref{fig-dm}(d).

\subsection{Small $t$}
In this subsection, we perform the calculation for small $t$ in detail. We will work out general $Q$ to further explain the second line of \refeq{eq-FS}. We take $|A|=|B|=L/2$ for convenience.

First, consider the second term $\mathcal{Z}_{bb}$. There is only one configuration at the lowest order of $1/d$: all spins should be equal to $b$. The number of fixed points $n_b=0$; the number of triangles equals $\frac{Lt}{2}$. Therefore, the partition function is:
\begin{equation}\label{eq-Zbbsmallt}
    \mathcal{Z}_{bb}\approx (1-p)^{\frac{QLt}{2}}.
\end{equation}
This also gives the Rényi-2 entropy of the whole system by \refeq{eq-S2whole}:
\begin{align}\label{eq-totalS}
 \overline{S_{2}}(t)&=-2\frac{\partial}{\partial Q}\log\mathcal{Z}_{bb}\approx (L\log\frac{1}{1-p})t.
\end{align}

\begin{figure}[h]
    \centering
    \includegraphics[width=\linewidth]{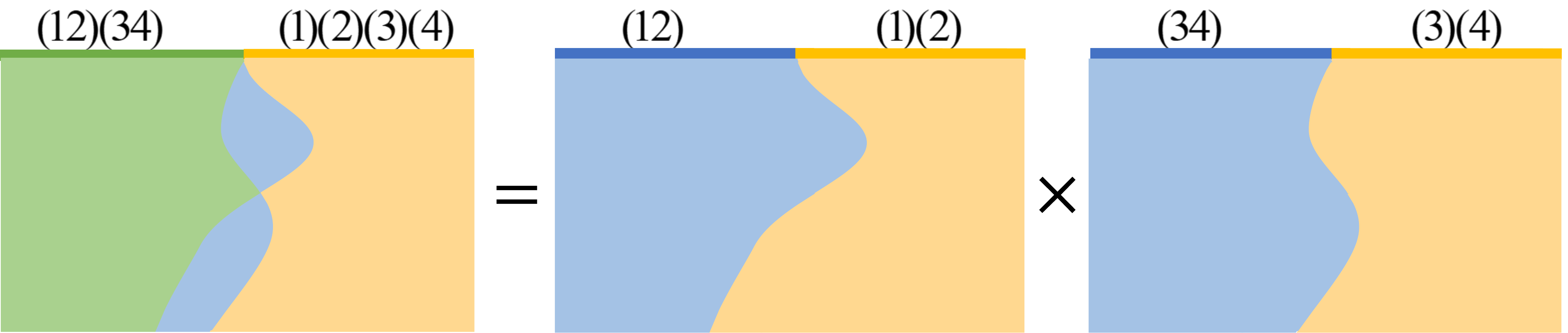}
    \caption{Domain wall configuration for $I_2$ at small $t$. Blue spins are $(12)$ or $(34)$; green spins are $(12)(34)$; yellow spins are $id$. Two regions are separated by a domain wall. The first figure shows a configuration with $Q=4$; it has two commuting domain walls. It can be decomposed as a product of two single domain wall configurations. The equation is valid even considering the $(1-p)^{Q-n}$ factors.}
    \label{fig-dmISt}
\end{figure}

Next, consider the first term $\mathcal{Z}_{ba}$, there must be $Q/2$ commuting domain walls starting from the intersection point and propagating vertically to the bottom. However, this time spins could have different numbers of fixed points. For example, in the middle figure of \reffig{fig-dmISt} we show a configuration for $Q=2$.  Spins on the right side of the domain wall are equal to $id$, hence they contribute 1 instead of $(1-p)^2$. 
Fortunately, if a configuration only contains $Q/2$ commuting domain walls, then one can decompose the configuration as a superposition of $Q/2$ configurations, each containing only one type of domain wall. The number of unfixed points exactly equals the summation of unfixed points for each:
\begin{equation}\label{eq-decom1}
    Q-n_a=\sum_i^{Q/2}(2-\tilde{n}_{a_i}).
\end{equation}
Here, we decompose $a\in S_Q$ as the product of $a_i$, where $a_i\in S_2^{(i)}=\{id,(2i+1,2i+2)\}$.  $\tilde{n}_{a_i}$ is the number of fixed points of $a_i$ defined in that $S_2^{(i)}$ (so $\tilde{n}_{id}=2$); see \reffig{fig-dmISt} for illustration. With \refeq{eq-decom1} and the discussions in \refsec{sec-Tweights} in mind, we have:
\begin{equation}\label{eq-decom2}
    \mathcal{Z}_{ba}(Q)\approx[\mathcal{Z}_{ba}(Q=2)]^{Q/2}.
\end{equation}
Therefore the calculation reduces to the case with only one domain wall as in \refeq{eq-FS}.

To calculate the entropic contribution, let us represent a domain wall by a vector $x=(x_1,x_2,\cdots,x_t)$, where $x_i=\pm 1$ if the $i$th step turns right or left (from our point of view). Then the number of $(1-p)^2$ factors equals:
\begin{equation}
\begin{aligned}
         &\frac{Lt}{4}-\frac{x_1}{2}+\frac{x_1+x_2}{2}+\cdots+\frac{x_1+x_2+\cdots+x_t}{2}\\
         =&\frac{Lt}{4}+\sum_{i=1}^t\frac{(t+1-i)x_i}{2}.   
\end{aligned}
\end{equation}
Therefore,
\begin{equation}\label{eq-Zabsmallt}
\begin{aligned}
    &\mathcal{Z}_{ab}(Q=2)=\mathcal{Z}_{ba}(Q=2)\\
    \approx&\frac{1}{d^t}\sum_{x}(1-p)^{\frac{Lt}{2}-\sum_{i=1}^t(t+1-i)x_i}\\
    =&\frac{2^t}{d^t}(1-p)^{\frac{Lt}{2}}\prod_{i=1}^t\frac{(1-p)^{-i}+(1-p)^{i}}{2}.
\end{aligned}
\end{equation}
Combining \cref{eq-Zbbsmallt,eq-Zabsmallt}) together, the Rényi-2 mutual information is given by:
\begin{equation}\label{eq-averageI2}
     \overline{I_{2}(A\colon B)}\approx (2\log\frac{d}{2})t-2\sum_{i=1}^t\log\frac{(1-p)^{-i}+(1-p)^{i}}{2}.
\end{equation}

We note that, all extensive (proportional to $L$) terms  in \cref{eq-Zbbsmallt,eq-Zabsmallt}) exactly cancel with each other, leaving us an $L$ independent expression. 
As a side note, the first term in this expression is not valid if $d=2$, since $d$ is not very large. In this case, we can replace $1/d$ with the exact vertical domain wall contribution $\frac{d}{d^2+1}$:
\begin{equation}\label{eq-newd}
  (2\log\frac{d}{2})t \to (2\log\frac{d^2+1}{2d})t.
\end{equation}

\subsection{Large $t$}

When $t$ is large enough, we expect the system to be almost maximally mixed:
\begin{equation}\label{eq-inftlimit}
    \rho(t\to\infty)=\frac{\mathbb{I}_L}{d^L}.
\end{equation}
Various partition functions in \cref{eq-FI,eq-FS} can be easily calculated in this limit. For example,
\begin{equation}\label{eq-Zbalimit0}
    \mathcal{Z}_{ba}=(\tr_A(\tr_B\frac{\mathbb{I}_L}{d^L})^2)^{\alpha}=\frac{1}{d^{|A|\alpha}}=\frac{1}{d^{|A|Q/2}}. 
\end{equation}
Similarly,
\begin{equation}\label{eq-Zbblimit0}
    \mathcal{Z}_{ab}=\frac{1}{d^{|B|Q/2}},~~\mathcal{Z}_{bb}=\frac{1}{d^{LQ/2}}.
\end{equation}
Therefore, $I_2$ vanishes as expected.

From the stat-mech model point of view, as discussed above, we need to consider domain walls separating the upper and lower boundaries as in Figs.~\ref{fig-dm}(c) and \ref{fig-dm}(d).. 
Interestingly, the sum over these fluctuating domain walls can be analytically carried out, under the following two restrictions: 
\begin{itemize}
    \item the number of domain walls is minimal (no ``bubbles");
    \item domain walls are \textit{horizontally directed}: projecting horizontally, the domain wall can never overlap.  
\end{itemize}
Figures~\ref{fig-dm}(c) and \ref{fig-dm}(d). are examples of configurations satisfying these restrictions.
The reason for imposing these restrictions is to keep the exponent of $1/d$ minimal: extra horizontal segments and bubbles will contribute more $1/d$ factors. 

The method for the summation is iterational (over $t$ or $L$) and  can be found in the Appendix \ref{app-larget}. 
Here we just mention that, as a sanity check, the partition functions indeed match the result with the maximally mixed final state.

\subsection{Thermalization timescale}

Comparing Eqs.~(\ref{eq-Zbbsmallt}) and (\ref{eq-Zabsmallt}) with the infinite $t$ result \cref{eq-Zbalimit0}, we see there is a competition between $1/d$ and $(1-p)^{t}$. Equating them yields the timescale
\begin{equation}\label{eq-timescale}
    t^*=O\left(\frac{\log d}{-\log(1-p)}\right).
\end{equation}

It is natural to interpret this timescale as the timescale of thermalization (in our case, trivialization). Below this timescale, vertical domain walls dominate, and the total system entropy \refeq{eq-totalS} still grows. After this timescale, horizontal domain walls dominate, total system entropy saturates, and mutual information vanishes. 

We emphasize that there is no $L$ (system size) dependence in $t^*$.
From the stat-mech perspective, this system-size-independent thermalization timescale reflects the fact that the fixed boundary condition has only short-range effects on the bulk spins. 
This is because the domain wall contribution $\sim\frac{1}{d^L}$ is boundary-like, but the polarization field contribution $\sim(1-p)^{Lt}$ is extensive. Equating them always give us a finite, $L$-independent timescale as in \refeq{eq-timescale}. 
Moreover, as discussed near \refeq{eq-Ising}, the physics resembles the 2D Ising model with a magnetic field. In this setting, the Gibbs state is well-known to be unique and short-range correlated for any nonzero magnetic field. The same is true for more general models with a polarizing field like the Potts model \cite{PhysRevB.24.1374,TSAI2009485}, which are relevant for higher $Q$.

\section{Area law}

The system-size-independent growth rate in \refeq{eq-averageI2} can be regarded as an open system analog of the small incremental theorem \cite{BravyiSIE,marien2016entanglement}. It can be rigorously proven for the von Neumann mutual information under general local quantum channel dynamics.  

Let us consider $I(A\colon B)$ for any bipartite system $AB$. There are two possibilities for each local quantum channels $\Phi$. If a local channel $\Phi$ acts inside $A$ or $B$, then $I(\Phi(A)\colon \Phi(B))\leq I(A\colon B)$ according to the data processing inequality/monotonicity of mutual information. If a local channel $\Phi$ acts on the boundary between $A$ and $B$, then $I(A\colon B)$ can only increase a constant. Indeed, denote the qubits at the boundary as $a$ and $b$ ($a\in A$ and $b\in B$), then after the action, we have:
\begin{equation}
    \begin{aligned}
    S'(A)&\leq S'(A\backslash a)+S'(a) = S(A\backslash a)+S'(a) \\
    &\leq S(A)+S(a)+S'(a),
    \end{aligned}
\end{equation}
since $\Phi$ does not touch $A\backslash a$ (the same for $B$), and
\begin{equation}
    S'(AB)\geq S(AB),
\end{equation}
since $\Phi$ is unital. Therefore,
\begin{equation}
    I'(A\colon B)\leq S(a)+S'(a)+S(b)+S'(b)+I(A\colon B).
\end{equation}
Since there is only one boundary unitary in each time slice, $I(A\colon B)$ grows at most linearly. The maximal slope is $O(\log d)$, which matches \refeq{eq-averageI2}.

We see that this almost-linear growth is valid not only at the level of trajectory average but also for each trajectory.

Moreover, recall that we have a system-size-independent timescale \refeq{eq-timescale} after which the system trivializes and the mutual information vanishes. Combined with the system-size-independent growth rate, it leads us to an \textit{area law}: the peak of mutual information must be bounded by the size of the bipartition boundary (hence ``area law") instead of any volume. 

\section{Probabilistic trace setup}
A depolarizing channel $\Phi(\rho)=(1-p)\rho+\frac{1}{d^2}\mathbb{I}_2$ can be regarded as a probabilistic mixture of an identity channel $\rho\mapsto\rho$ and a ``trace channel" $\rho\mapsto \frac{1}{d^2}\mathbb{I}_2$. Namely, with probability $p$, the two qubits undergo some noise and become completely trivialized; with probability $(1-p)$, nothing happens. In reality, we usually do not know whether the noise happens or not, so we need to use the depolarizing channel to describe the noisy process.

However, the scenario of knowing whether the noise happens has its advantage. Namely, if we further restrict all unitaries to Clifford unitaries, the whole quantum process will be classically simulatable: since both  trace channels and Clifford unitaries send stabilizer states to stabilizer states, and the initial state (pure product state of $\ket{0}$) is a stabilizer state, we can effectively simulate the whole quantum process using the stabilizer formalism \cite{gottesman1997stabilizer}. 

\subsection{Setup}

\begin{figure}
\centering
\includegraphics[width=0.6\linewidth]{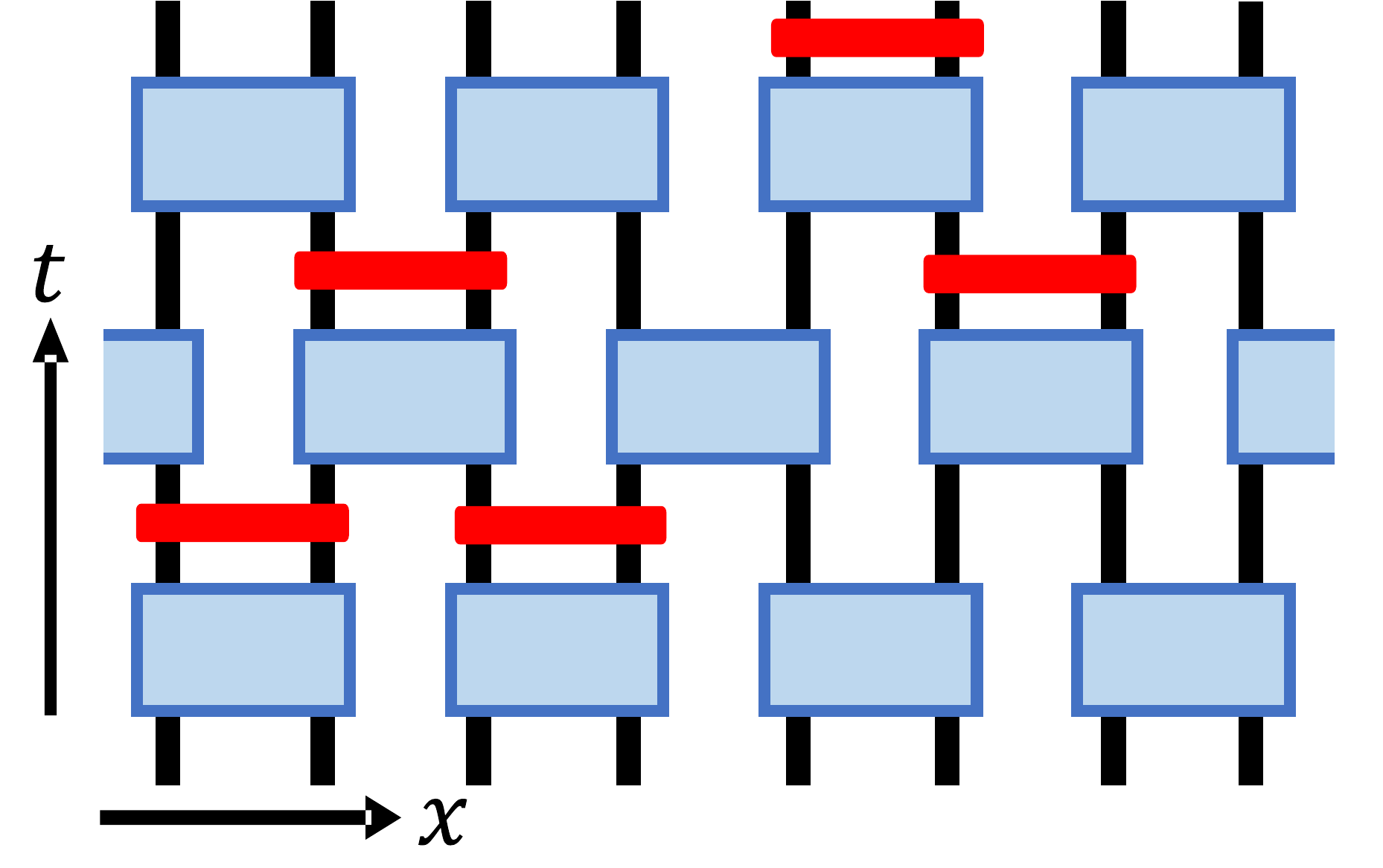}
\caption{The probabilistic trace setup. Trace channels (red strips) are applied with probability $p$ after random Haar or Clifford unitaries. }\label{fig-probcircuit} 
\end{figure}

The ``probabilistic trace setup" is shown in \reffig{fig-probcircuit}. We still have independent random unitaries $U$ represented as blue blocks.
On top of each unitary, we apply with probability $p$ a  trace channel $\Phi_T(\rho)=\frac{\tr(\rho)}{d^2}\mathbb{I}_2$. 
Conceptually, this setup also capture the dynamics of open quantum systems with strength $O(p)$ depolarization.

Note that, different from the original random channel setup, here each quantum trajectory/circuit is realized by fixing the unitaries \textit{as well as} the presence/nonpresence of each trace channel. Individual trajectory $\rho(t)$ does not contains $p$. For each trajectory $\rho(t)$, we still consider its bipartite mutual information $I_2(t)$. Similar to \refeq{eq-Emutual}, we again focus on the average over circuit realizations. This time we need to average over both unitaries and the presence/nonpresence of traces:
\begin{equation}
    \mathbb{E}[I_2(\rho_t)]=\mathbb{E}_U\mathbb{E}_T[I_2(\rho_t)].
\end{equation}
The parameter $p$ appears only at the level of averaging.

A trajectory $\rho^{(1)}$ in the previous random channel setup can be regarded as an averaging over trajectories $\rho^{(2)}$ in the probabilistic trace setup (fix all the unitaries and only average over the appearance of traces):
\begin{equation}
   \rho^{(1)}=\mathbb{E}_T(\rho^{(2)}).   
\end{equation}
However, this does not imply any equation between $\mathbb{E}[I_2(\rho^{(1)})]$ [the quantity considered in \refeq{eq-Emutual}] and $\mathbb{E}[I_2(\rho^{(2)})]$ (the quantity to be considered in the new setting), since $I_2$ is not a linear function and does not commute with $\mathbb{E}_T$. Nevertheless, for any convex function $\mathcal{M}$, we have:
\begin{equation}\label{eq-convex}
    \mathbb{E}_U\mathcal{M}(\rho^{(1)})\leq \mathbb{E}_U\mathbb{E}_T[\mathcal{M}(\rho^{(2)})]=\mathbb{E}[\mathcal{M}(\rho^{(2)})]
\end{equation}

\subsection{Stat-mech mapping and Area law}

The new setup can also be mapped into a stat-mech model, see Appendix \ref{sec-probtracemap} for details. Here, we just list some relevant triangle weights in \reftab{tab-weights-prob}. In this subsection, the unitaries are chosen from the Haar ensemble.

\begin{table}[h]
    \centering
    \begin{tabular}{c||c|c|c|c|c|c|c}
        \hline\hline
         $J_p(a,b,c)$ &$\Tooo$ &$\Tsos$ &$\Toss$ &$\Todd$ & $\Tsds$& $\Tdss$&$\Tdod$ \\
         \hline
         $a\neq id$& $1-p$ & $\frac{1-p}{d}$ & 0 & 0 &0 &$\frac{1-p}{d^2}$ &$\frac{1-p}{d^2}$  \\
         \hline
        $a=id$&1 & $1/d$ & $\frac{p}{d^2}$& $\frac{p}{d^4}$&$\frac{p}{d^3}$ &$\frac{1}{d^2}$ &$\frac{1}{d^2}$  \\
        \hline
        $p=0$& 1 & $1/d$ & 0 & 0 & 0 &$\frac{1}{d^2}$ &$\frac{1}{d^2}$  \\
         \hline\hline
    \end{tabular}
    \caption{Triangle weights $J_p$ for the statistical model. Here $a$ is the spin of the downward vertex.}\label{tab-weights-prob}
\end{table}


We see that the physical effects of the noise are still similar as before:
\begin{itemize}
    \item Polarization field. If $p>0$, then the bottom spins are more probable to be $id$, due to a relative weight of 1 versus $(1-p)$. 
    \item Horizontal domain wall. If $p>0$, then domain walls can propagate horizontally. This is only possible when the bottom spin equals $id$, again with an extra penalty of $1/d$ and a factor (here it is $p$) to forbid horizontal domain walls in the  $p=0$ case.
\end{itemize}

Up to $O(1/d)$, the triangle weights in \reftab{tab-weights-prob} can be equivalently summarized as:
\begin{equation}
    E=-\log d\sum_{\braket{ij}}(\delta_{\sigma_i\sigma_j}-1)-\log\frac{1}{1-p}\sum_i(\delta_{\sigma_i}-1),
\end{equation}
which is the $Q!$-state Potts model with magnetic field on a (rotated) two-dimensional square lattice. For $Q=2$ it goes back to the Ising model \refeq{eq-Ising}. Moreover, for $Q=2$, if we define $p'$ by
\begin{equation}
    (1-p')^2=1-p,
\end{equation}
then the weights have the same form as in the previous setup \refeq{eq-Tgeneral} (in terms of $p'$) even up to the order of $O(1/d^2)$.

Therefore, two models should share very similar properties. In particular, the system should thermalize in a system-size-independent time, and the mutual information peak value should obey an area law.

However, the rigorous analytical treatment in this case is more complicated. If $Q>2$, then triangle weights no longer factorize as \refeq{eq-Tgeneral}. This means interactions between replicas are no longer negligible. For example, denote $a=(1,2)(3,4)$, $b=(1,2)$, then:
\begin{equation}
\begin{aligned}
    J_p(id,a,a)=\frac{p}{d^4}&\neq J_p(id,b,b)J_p(id,b,b)=\frac{p^2}{d^4}\\
    &\neq J_p(b,a,a)J_p(b,a,a)=0.
\end{aligned}
\end{equation}
As another example of the difference, the counterpart of \refeq{eq-Zbbsmallt} will be $(1-p)^{Lt/2}$ for any $Q$.

\subsection{Clifford numerics}
As discussed at the beginning of this section, the main motivation for considering this setting is the ability of scalable classical simulation if the unitaries are Clifford. In this subsection, we discuss the results of Clifford numerics. Besides the mutual information, we also simulated the entanglement negativity, see the Appendix for details. 

First, we comment on some properties of stabilizer states. 
\begin{itemize}
    \item Stabilizer states (pure or mixed) always have flat entanglement spectra, so the Rényi entropy does not depend on the index $n$. The statement also holds for Rényi mutual information and negativity. 
    \item The entanglement structure of stabilizer states is very clear due to a structure theorem \cite{bravyi2006ghz}: (1) operator entanglement entropy equals the mutual information, which counts the amount of both classical and quantum correlations; (2) negativity counts the quantum part of the correlations.
\end{itemize}



\begin{figure}
\centering
\subfloat[]{\includegraphics[width=0.25\textwidth]{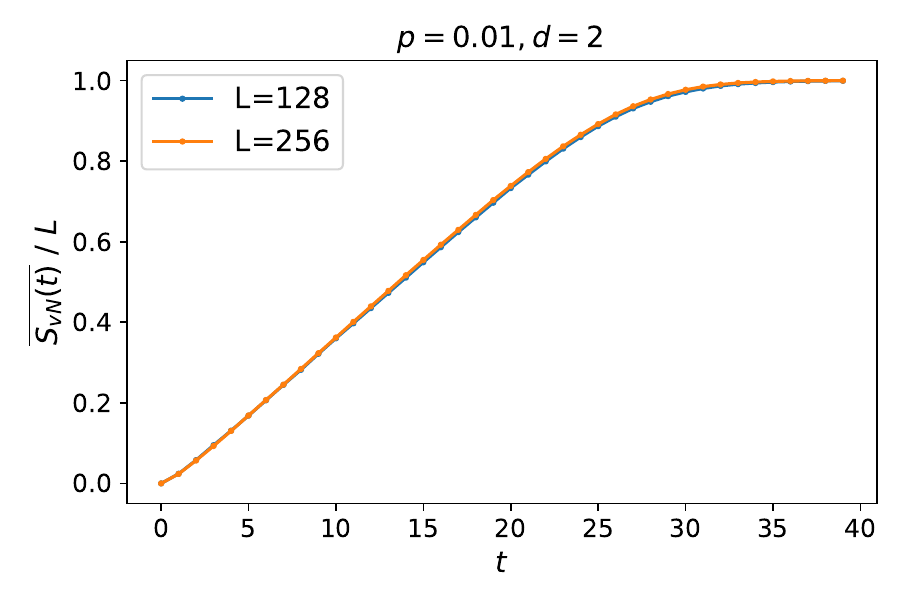}}
\subfloat[]{\includegraphics[width=0.25\textwidth]{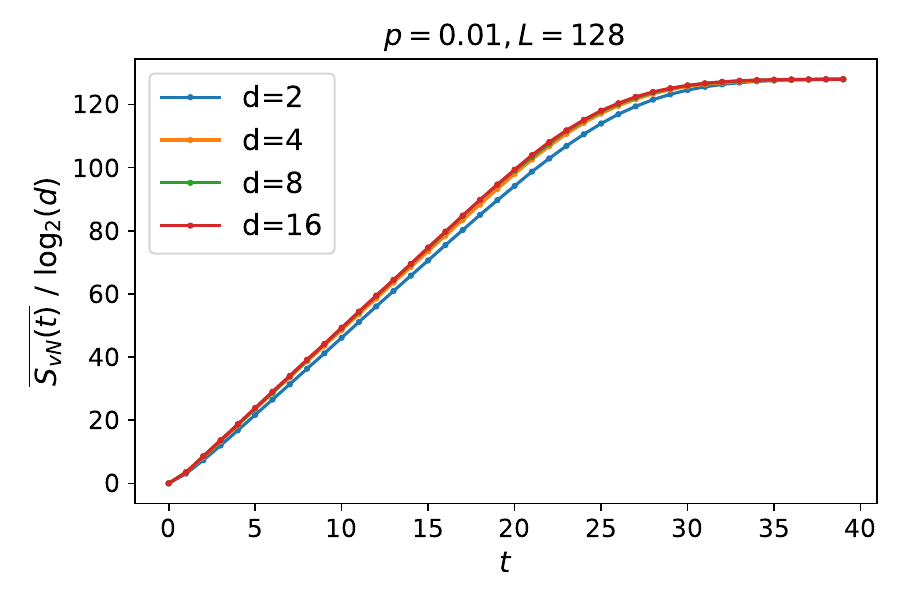}}
\caption{
Simulated state's  von-Neumann entropy $S_{\text{vN}}(t)$ for various different (a) system size $L$, (b) onsite dimension $d$.}\label{data-vN}
\end{figure}

\begin{figure*}
\centering
\subfloat[]{{\includegraphics[width=0.325\textwidth]{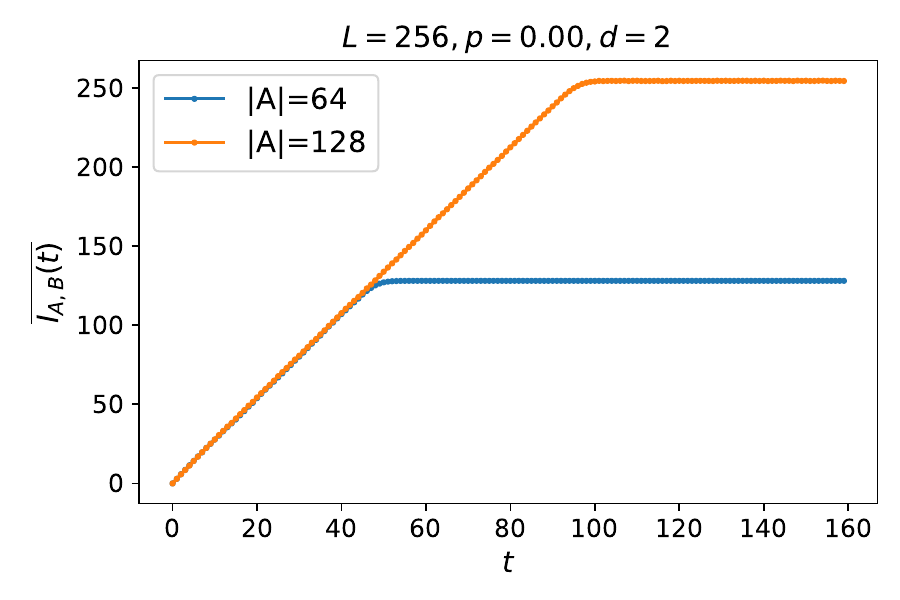}}}
\subfloat[]{\includegraphics[width=0.325\textwidth]{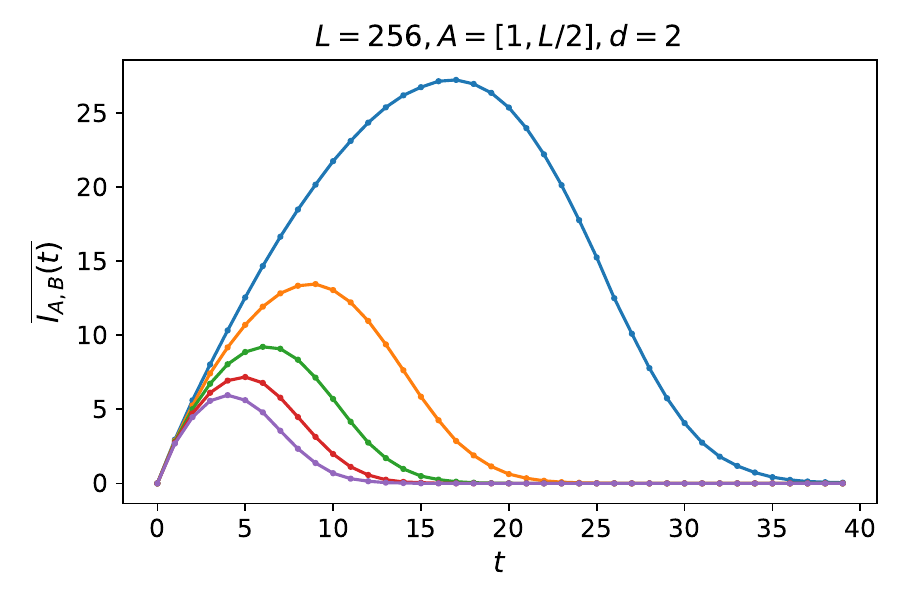}}
\subfloat[]{\includegraphics[width=0.325\textwidth]{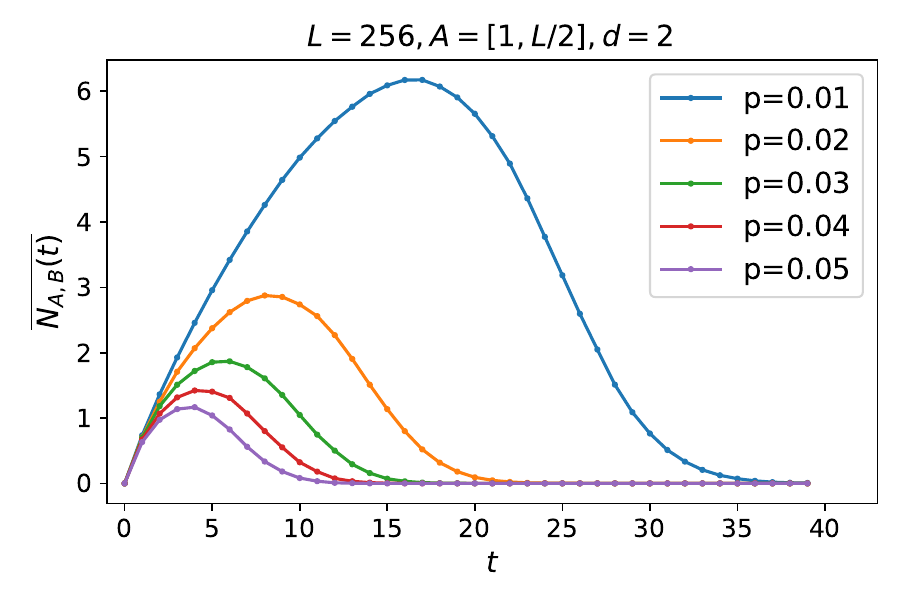}}
\\[-2ex]\caption{(a) Mutual information for the $p=0$ unitary-only case. 
(b) Mutual information for the nonzero $p$ values. (c) Logarithmic entanglement negativity between $A$ and $B$ for various nonzero $p$ values.}\label{data-mutual}
\end{figure*}

\begin{figure}
\centering
\subfloat[]{\includegraphics[width=0.25\textwidth]{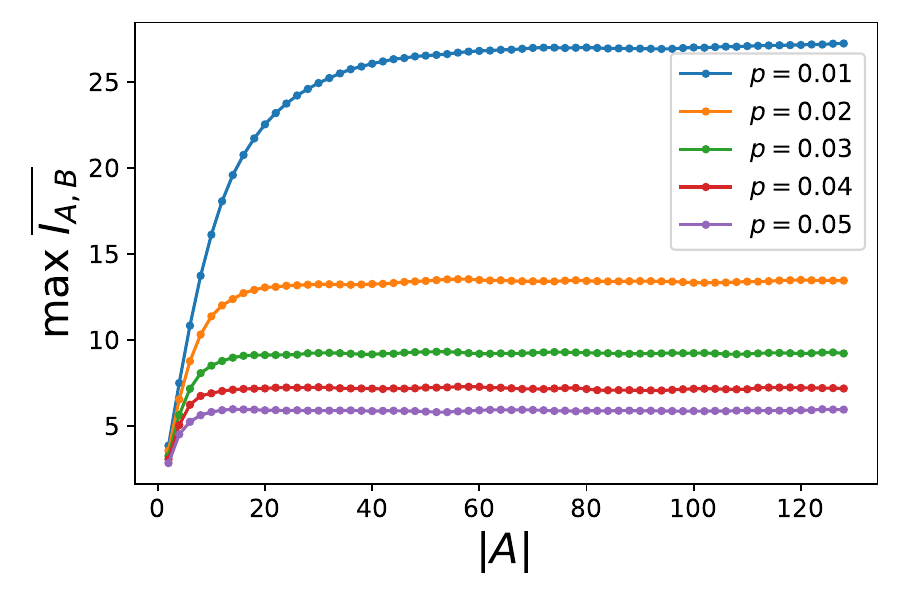}}
\subfloat[]{\includegraphics[width=0.25\textwidth]{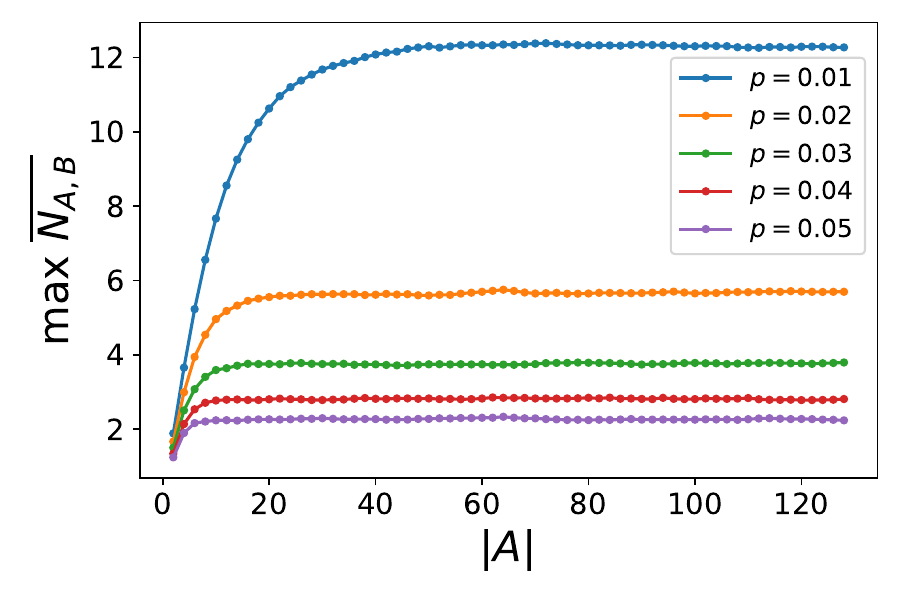}}
\caption{
(a)  Maximum mutual information and (b) maximum log-negativity during the dynamics for various $p$ and $|A|$. $L=256$ in simulations above.}\label{fig-data} 
\end{figure}


In Figs. \ref{data-vN}(a) and \ref{data-vN}(b), we show the simulation for the total system entropy $S(t)$. We see that it grows and saturates to a volume law value in $O(1)$ time, and takes the form:
\begin{equation}
S(t) \approx (L\log d) f(p, t)
\end{equation}
For late time, $f(p, t)$ curves toward its limiting value $f(p, \infty) = 1$, so that $S(t)$ converges to its limiting value $L\log d$, which is the von Neumann entropy of a maximally mixed state.
The $\log d$ dependence of the slope is different from the one predicted in \refeq{eq-totalS} for the previous setup. We believe it is a subtle difference between two architectures, see Eqs.~\eqref{eq-toyRenyin} and~\eqref{eq-toy2averageS} for the difference in a 0-dimensional toy model.

In \reffig{data-mutual} we show the simulated results for bipartite entanglement measures (mutual information and log-negativity). As a comparison and benchmark, for $p=0$  unitary-only case \reffig{data-mutual}(a), the mutual information grows and saturates to a volume law (proportional to $|A|$) plateau. Figure \ref{data-mutual}(b) shows how $I(A,B)$ grows and reaches some peak value and then quickly decays to zero if $p\neq 0$. Figure \ref{data-mutual}(c)  is the result for the log-negativity. It behaves similarly as $I(A\colon B)$, indicating that the classical and the quantum parts of the correlation have qualitatively similar behavior.

To verify the area law, in \reffig{fig-data}, we show the peak values of mutual information and negativity for different $p$ and subsystem sizes $|A|$. It is clear that the peaks for both quantities are $|A|$-independent and hence obey an area law. In fact, the entire dynamics for two different partitions $|A|=L/4, L/2$ are nearly identical, which follow a profile as curves in Figs~\ref{data-mutual}(b) and~\ref{data-mutual}(c).

We note that area law in the probabilistic trace setup implies area law in the previous random channel setup for any convex entanglement measure $\mathcal{M}$ due to \refeq{eq-convex}. While neither the mutual information nor the log-negativity (what we plotted) is convex, the negativity itself is convex. Moreover, for stabilizer states, the structure theorem \cite{bravyi2006ghz} implies that the log-negativity equals the squashed entanglement \cite{tucci2002entanglement}, the latter being a nice convex entanglement measure for general states. 

\section{Discussions} 

Using both an analytic mapping to a spin model and large-scale Clifford simulations, we showed that systems evolving under random unitaries and depolarization channels thermalize at an $L$-independent timescale \refeq{eq-timescale}. Correspondingly, various entanglement measures have peaks obeying area laws.   This implies, in addition to \refcite{MPOnumerics}, that matrix product operator simulations of such noisy 1D dynamics are in principle efficient, although in practice the required bond dimension may still be large for small noise strength $p$. This indicates that noisy random circuit sampling is not likely to provide a quantum computational advantage.

An immediate question is whether the area law still holds in higher dimensions.
In Appendix \ref{subsec:2d}, we also consider a 2D system evolving under random Clifford gates and depolarization, and the numerical results still suggest the area law. A rigorous analytical treatment in higher dimensions is worth exploring.

\begin{figure*}
\centering
\subfloat[]{\includegraphics[width=0.48\textwidth,valign=b]{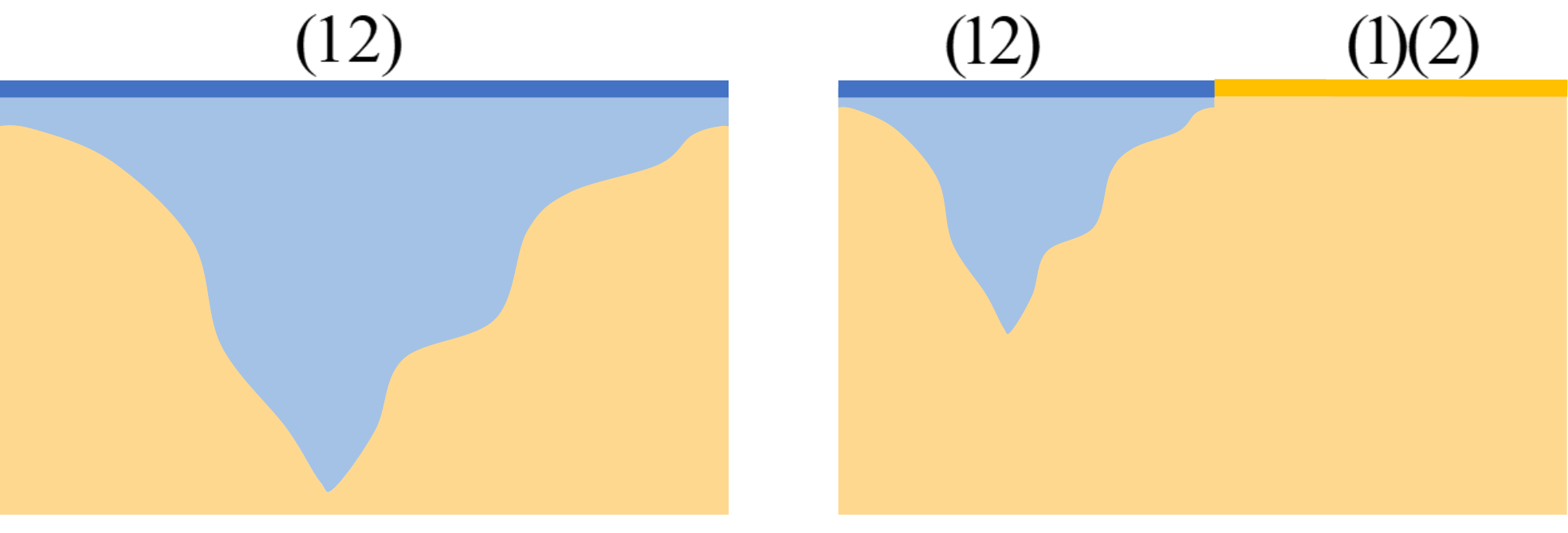}}
\subfloat[]{\includegraphics[width=0.24\textwidth,valign=b]{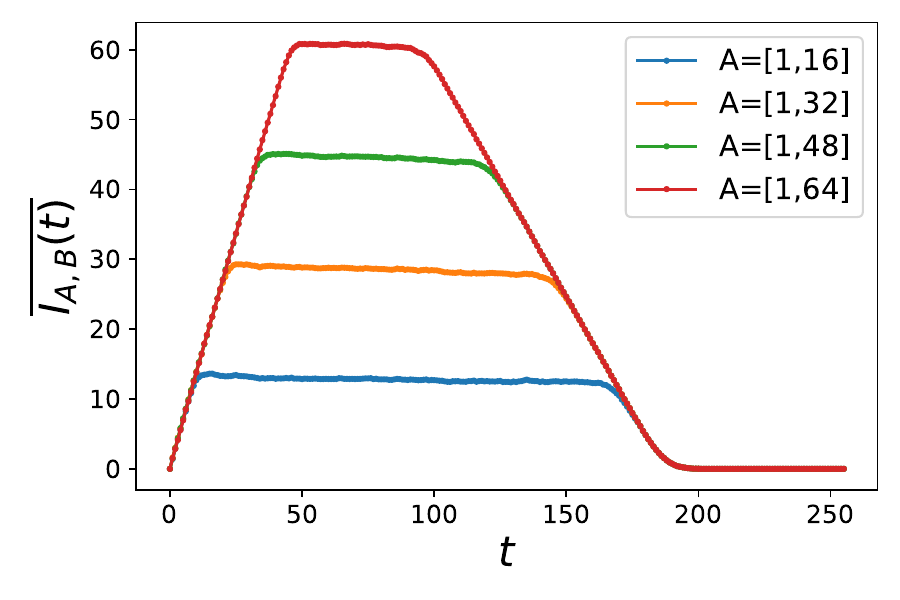}}
\subfloat[]{\includegraphics[width=0.24\textwidth,valign=b]{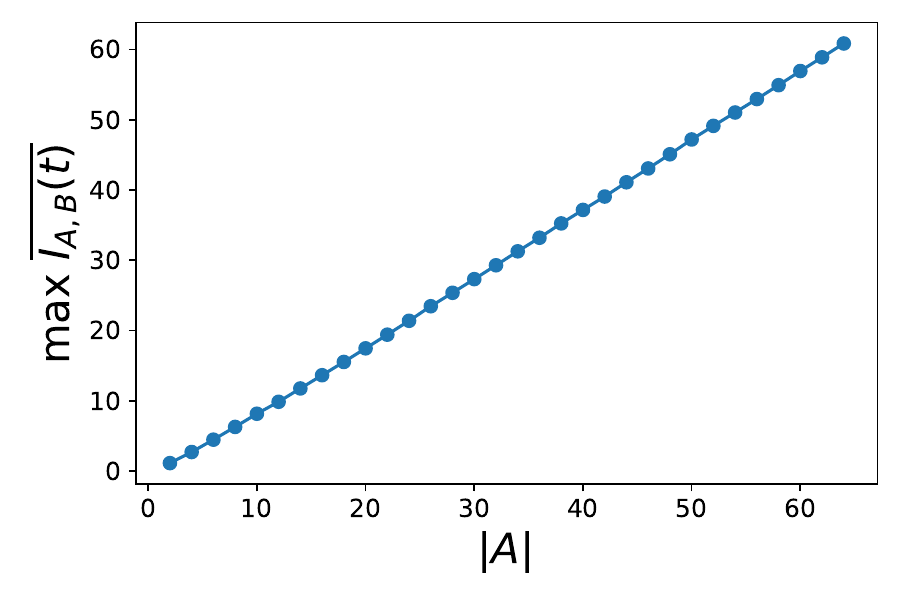}}
\caption{Analysis and numerics for boundary-only depolarization setting. (a) Domain wall configurations of the corresponding spin model. (b) Dynamics of the mutual information for various $|A|$ in the Clifford version.  (c) Linear regression between the peak value and $|A|$. $L=128$ in simulations above.}
    \label{fig-bdonly}
\end{figure*}

The $L$-independent timescale originates from the extensiveness of the depolarization. In contrast, let us consider a model where depolarizations only apply at the boundary. 
In this situation, we can still perform similar mapping, resulting in a classical spin model. In the spin model, triangle weights in the bulk are the same as the $p=0$ unitary-only case and only vertical domain walls are allowed. The only difference happens at the boundary, where horizontal domain walls are allowed. 
To reach thermalization such that spins deep in the bulk are $id$, domain walls should look like \reffig{fig-bdonly}(a). The timescale is at least $O(L)$ to allow such configurations. This indicates that the thermalization timescale will be $O(L)$ for the boundary-only depolarization  model.
In \reffig{fig-bdonly}(b), we show the numerical results for a Clifford random circuit with boundary-only trace channels applied definitely ($p=1$). The peak value clearly obeys a volume law, as verified in \reffig{fig-bdonly}(c).

The case of boundary-only depolarization may be related to a contiguous subsystem of a closed system evolving under random unitaries. Effectively, this subsystem is coupled to its ``environment'' (the complement) through its boundary, which acts as a  boundary-only depolarization. Reference~\cite{wang2019barrier} found that the bipartite operator entanglement of such a subsystem exhibits a volume law peak during the dynamics, consistent with our analysis and numerics.

\begin{acknowledgments} 
We thank Tarun Grover, Yaodong Li, and Beni Yoshida for helpful discussions. This work was supported by Perimeter Institute, NSERC RGPIN-2018-04380, and Compute Canada.  Research at Perimeter Institute is supported in part by the Government of Canada through the Department of Innovation, Science and Economic Development and by the Province of Ontario through the Ministry of Colleges and Universities.
\end{acknowledgments}

\bibliography{main.bib}

\appendix
\section{More on the Stat-mech Model}
\subsection{Correlation and entanglement measures}

In the main text, we considered the Rényi-2 mutual information. We could also consider the Rényi-$n$ mutual information:
\begin{equation}\label{eq-renyinmutual}
\begin{aligned}
    I_n(A\colon B)=&S_{n,A}+S_{n,B}-S_{n,AB}\\
    =&\frac{1}{1-n}(\log\tr\rho_A^n-\log\tr\rho_B^n+\log\tr\rho_{AB}^n),
\end{aligned}
\end{equation}
where $S_{n,*}$ is the Rényi-$n$ entropy of the reduced density matrices $\rho_*$ where $*$ stands for (sub)systems $A$, $B$ and $AB$.

The operator entanglement entropy for a (pure or mixed) state is the entanglement entropy for a corresponding normalized operator state. More precisely, a density matrix $\rho$ on a bipartite system $AB$ can be regarded as a pure state $\ket{\rho}$ in $H_{AB}\otimes H^*_{AB}$, where $H^*_{AB}$ is the dual space of $H_{AB}$. This state in general should be normalized by $\norm{\rho}^2=\braket{\rho|\rho}=\tr\rho^2$.
The bipartite operator entanglement entropy is then:
\begin{dmath}\label{eq-renyiopEE}
 S_{n,A,B}^{\text{op}}=\frac{1}{1-n}\log\widetilde{\tr}_A\left(\widetilde{\tr}_B\frac{\ket{\rho}\bra{\rho}}{\norm{\rho}^2}\right)^n=\frac{1}{1-n}\left\{\log\widetilde\tr_A\left[\widetilde\tr_B(\ket{\rho}\bra{\rho})\right]^n-\log(\tr\rho^2)^n\right\}.
\end{dmath}
Here, we use $\widetilde\tr$ for traces of density matrices  on the doubled Hilbert space $H_{AB}\otimes H^*_{AB}$.

The operator entanglement entropy measures the complexity to represent the density matrix as a matrix product operator (MPO), similar to the usual entanglement entropy measuring the complexity to represent the wave-function as a matrix product state (MPS). Like mutual information, it also measures both classical and quantum correlations.

The entanglement negativity is a useful quantity to measure the quantum entanglement for a bipartite system. The logarithmic negativity is defined as:
\begin{equation}
    N_{A,B}=\log\norm{\rho^{\Gamma_A}}_1,
\end{equation}
where $\rho^{\Gamma_A}$ is the partial transpose on $B$, $\norm{\cdot}_1$ is the trace norm---sum of all singular values. Its Rényi generalization is defined as:
\begin{equation}\label{eq-renyinegtivity}
    N_{n,A,B}=-\log\frac{\norm{\rho^{\Gamma_A}}_n}{\tr\rho^n}=-\log\tr\left[(\rho^{\Gamma_A})^n\right]+\log\tr\rho^n,
\end{equation}
where $\norm{\cdot}_1$ is the $L^n$ norm. The second equation holds if $n$ is an even integer.


\subsection{More on boundary conditions}

For $I_n(A\colon B)$, if the replica number $\alpha=1$, then the first term amounts to fixing spins above $B$ region to $a=id_n$ and fixing spins above $A$ region to $b=(1,2,\cdots,n)$; the second term is similar; the third term amounts to fixing spins to $b$ everywhere. With $\alpha\geq 2$, we need to repeat the above pattern $\alpha$ times, so $a=id_Q$, $b=(1,2,\cdots,n)(n+1,n+2,\cdots,2n)\cdots(\cdots,Q)$, here $Q=n\alpha$.
At the end, \refeq{eq-renyinmutual} becomes
\begin{equation}\label{eq-FIn}
 \overline{I_{n}(A\colon B)}
 =\frac{n}{1-n}\frac{\partial}{\partial Q}\left(\log\mathcal{Z}_{ba}+\log\mathcal{Z}_{ab}-\log\mathcal{Z}_{bb}\right).
\end{equation}
where the subscripts  indicate different boundary conditions in the partition function $\mathcal{Z}$.

For $S_{n,A,B}^{\text{op}}$ and $\alpha=1$, the first term amounts to fixing spins above $A$ region to $c=(2,3)(4,5)\cdots(2n,1)$ and fixing spins above $B$ region to $d=(1,2)(3,4)\cdots(2n-1,2n)$, the second term amounts to fixing spins to $d$ everywhere. With $\alpha\geq 2$, $c$ becomes $(2,3)(4,5)\cdots(2n,1)(2n+2,2n+3)\cdots(4n,2n+1)\cdots(Q,Q-2n+1)$, $d$ becomes $(1,2)(3,4)\cdots(Q-1,Q)$, here $Q=2n\alpha$. Equation \eqref{eq-renyiopEE} becomes
\begin{equation}\label{eq-appFS}
 \overline{S_{n,A,B}^{\text{op}}}
 =\frac{2n}{1-n}\frac{\partial}{\partial Q}\left(\log\mathcal{Z}_{cd}-\log\mathcal{Z}_{dd}\right).
\end{equation}

For $N_{n,A,B}$ and $\alpha=1$, the first term amounts to fixing spins above $A$ to $e=(n,n-1\cdots,1)$ and fixing spins above $B$ to $b=(1,2,\cdots,n)$, the second term amounts to fixing spins to $b$ everywhere. With $\alpha\geq 2$, $e$ becomes $(n,n-1\cdots,1)(2n,2n-1\cdots,n+1)\cdots(Q,Q-1\cdots,Q-n+1)$, $b$ is the same as in $I_n(A\colon B)$. Here $Q=n\alpha$. \refeq{eq-renyiopEE} becomes
\begin{equation}\label{eq-FN}
 \overline{N_{n,A,B}}
 =n\frac{\partial}{\partial Q}(\log\mathcal{Z}_{eb}-\log\mathcal{Z}_{bb}).
\end{equation}

\section{More Large-$d$ Analysis}
\subsection{small $t$ calculation of $S_2^{\text{op}}$}

The calculation of $S_2^{\text{op}}$ is actually easier than the mutual information. 

\begin{figure}[h]
    \centering
    \includegraphics[width=\linewidth]{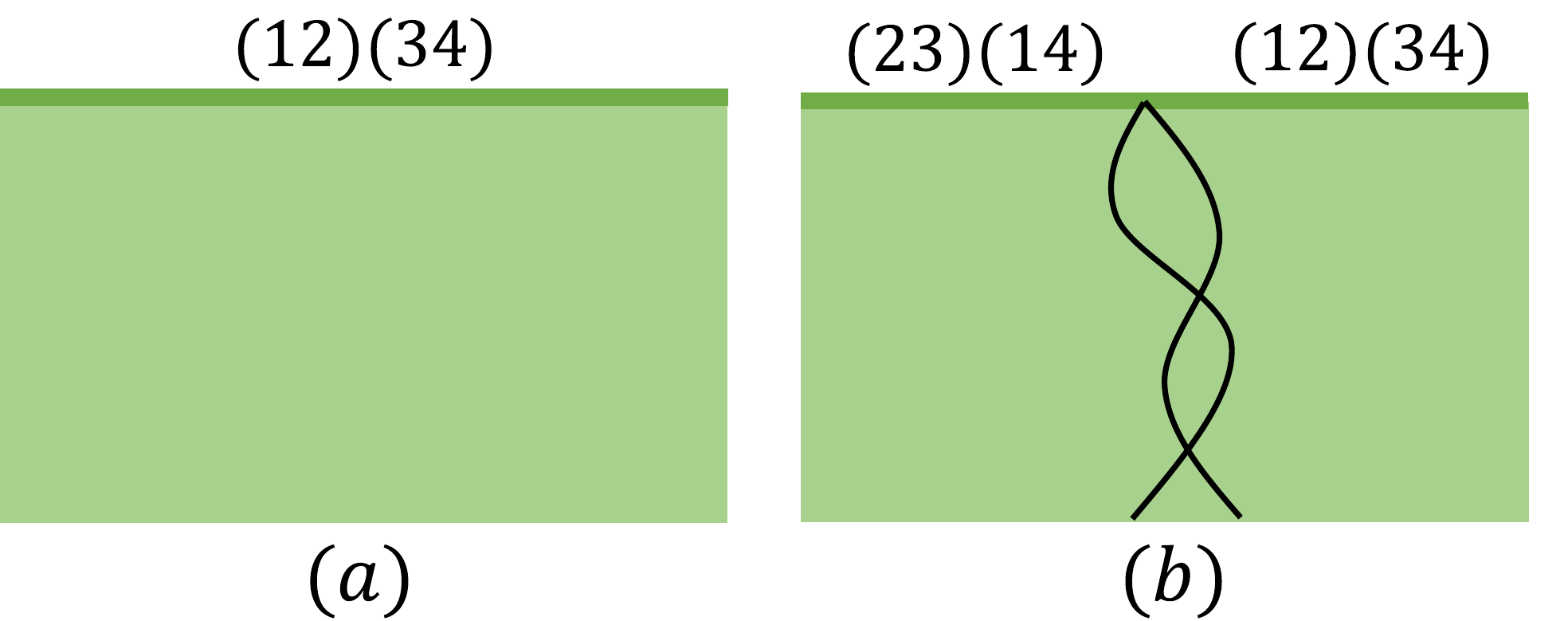}
    \caption{Domain wall configuration for $S^{\text{op}}$ at small $t$ ($Q=4$ for illustration). The thick line on the upper boundary indicates the fixed boundary condition. In panel (a), all spins are $b=(12)(34)$. In panel (b), spins can be different, hence the domain wall. In this case, all spins have the same number of fixed points (=0), so they are colored with the same color.}
    \label{fig-dmSopSt}
\end{figure}

For the first term $\mathcal{Z}_{cd}$ in \refeq{eq-appFS}, since $c^{-1}d=(1,3)(2,4)\cdots$, we know there must be $Q/2$ commuting domain walls starting from the intersection point. To get the lowest order in $1/d$, the domain walls need to propagate vertically to the bottom (because horizontal domain walls cost some extra factors of $1/d^2$); see \reffig{fig-dmSopSt}(b). Therefore each domain wall contributes a weight $1/d^{t}$. Each domain wall also has an entropic contribution $2^{t}$: it can go left or right at each step. Moreover, it is easy to check that all spins have no fixed points no matter how these domain walls locate, hence each triangle also contributes a $(1-p)^Q$. Therefore,
\begin{equation}\label{eq-appZcd}
    \mathcal{Z}_{cd}\approx (1-p)^{\frac{QLt}{2}}(\frac{2^t}{d^t})^{\frac{Q}{2}}.
\end{equation}

The second term $\mathcal{Z}_{dd}$ is similar to \refeq{eq-Zbbsmallt} in the main text. There is only one configuration with the lowest order of $1/d$: all spins should be equal to $d$ [see \reffig{fig-dmSopSt}(a)]. The number of fixed points $n_d=0$, the number of triangles is $\frac{Lt}{2}$, so:
\begin{equation}\label{eq-Zddsmallt}
    \mathcal{Z}_{dd}\approx (1-p)^{\frac{QLt}{2}}.
\end{equation}

Combining it with \refeq{eq-appZcd}, we get:
\begin{equation}
     \overline{S_{2,A,B}^{\text{op}}}(t)=-4\frac{\partial}{\partial Q}(\log\mathcal{Z}_{cd}-\log\mathcal{Z}_{dd})\approx (2\log\frac{d}{2})t.\label{eq-renyiopEE2}
\end{equation}

\subsection{Large $t$ calculation}\label{app-larget}
In this subsection, we show that the summing over horizontal domain wall configurations can be exactly carried out in the statistical mechanics model, and the results match with the limits at \refeq{eq-inftlimit}. We eventually need to calculate the partition functions for arbitrary layer numbers. However, the logic in \cref{eq-decom1,eq-decom2} still applies here, so we only need to focus on $Q=2$.

As discussed in the main text, we sum over horizontally directed configurations. 
The relevant graphical rules are summarized as follows: 
\begin{itemize}
\item vertical-horizontal segment \rotatebox[origin=c]{20}{$\langle$} contributes $1/d$ (boundary contribution);
\item horizontal-horizontal segment \rotatebox[origin=c]{-90}{$\langle$} contributes $\frac{1-(1-p)^2}{d^2}$ (boundary contribution); 
\item each triangle above the domain wall contributes $(1-p)^2$ (area contribution). 
\end{itemize}


\begin{figure}[h]
    \centering
    \includegraphics[width=\linewidth]{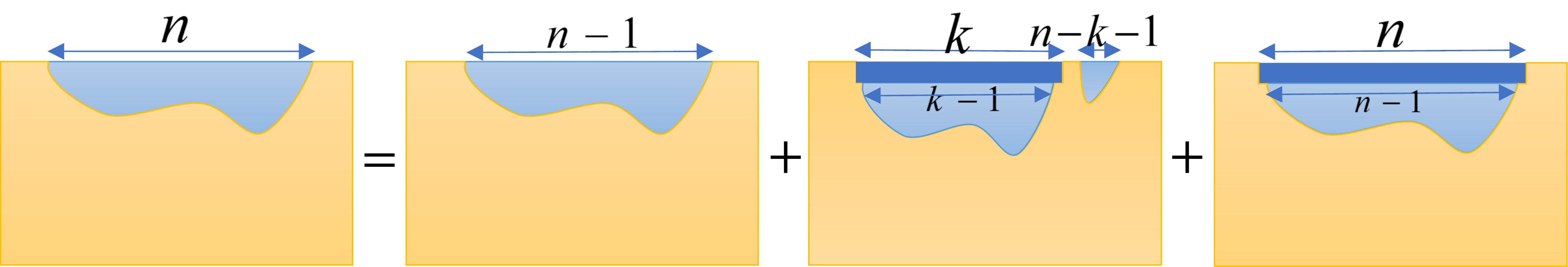}
    \caption{Iteration relations used in the calculation. Each rectangular represents a summation of some configurations. Dark blue means those spins are guaranteed to be $(12)$---in other words, the domain wall is at least 1 step away from the upper boundary; light blue means the domain wall can sometimes attached to the upper boundary.}
    \label{fig-iteration-zn}
\end{figure}
Let us first consider the ``hanging" configurations such that the distance between endpoints is $n$ (the domain wall is necessarily vertical at the endpoints). Denote the summation of these configurations as $u_n$. As shown in \reffig{fig-iteration-zn}, the domain wall can either go right at the first step, or firstly go down then return to the boundary at position $k~(1\leq k\leq n-1)$ then go right one more step, or firstly go down then return for the first time at position $n$. Therefore:
\begin{dmath}\label{eq-iteun}
    u_n=\frac{1-(1-p)^2}{d^2}u_{n-1}+\sum_{k=1}^{n-1}(1-p)^{2k}\frac{1}{d}u_{k-1}\frac{1}{d}\frac{1-(1-p)^2}{d^2}u_{n-1-k}+(1-p)^{2n}\frac{1}{d}u_{n-1}\frac{1}{d}.
\end{dmath}
With this iteration relation and the initial condition $u_0=1$, one can easily verify that
\begin{equation}
    u_n=\frac{1}{d^{2n}}.
\end{equation}

To calculate $\mathcal{Z}_{ba}$, denote $v_n$ to be the summation of configurations where domain walls' right endpoints are on the right boundary ($n$ is the distance between the left endpoint and the right boundary). Due to the (left and right) boundary conditions, there are actually two types of $v$, depending on whether the total number of layers is even or odd. We use $v$ and $v'$ to distinguish them. Similarly to \refeq{eq-iteun}, we have:
\begin{dmath}\label{eq-itevn}
     v_n=\frac{1-(1-p)^2}{d^2}v_{n-1}+\sum_{k=1}^{n-1}(1-p)^{2k}\frac{1}{d}u_{k-1}\frac{1}{d}\frac{1-(1-p)^2}{d^2}v_{n-1-k}+(1-p)^{2n}\frac{1}{d}v'_{n},
\end{dmath}
and
\begin{dmath}
v'_n=\frac{1-(1-p)^2}{d^2}v'_{n-1}+\sum_{k=1}^{n-1}(1-p)^{2k}\frac{1}{d}u_{k-1}\frac{1}{d}\frac{1-(1-p)^2}{d^2}v'_{n-1-k}+(1-p)^{2n}\frac{1}{d}v_{n-1}.
\end{dmath}
With this iteration relation and the initial condition $v'_0=1$, one can verify that:
\begin{equation}\label{eq-vtildev}
    v_n=\frac{1}{d^{2n+1}},~ v'_n=\frac{1}{d^{2n}}.
\end{equation}
(Less rigorously, one can regard this result as the limit of \refeq{eq-iteun} by taking the right endpoint to the right boundary.)
Therefore,
\begin{equation}
    \mathcal{Z}_{ba}=\frac{1}{d^{|A|}}.
\end{equation}
It is consistent with \refeq{eq-Zbalimit0}.

\begin{figure}[h]
    \centering
    \includegraphics[width=0.45\textwidth]{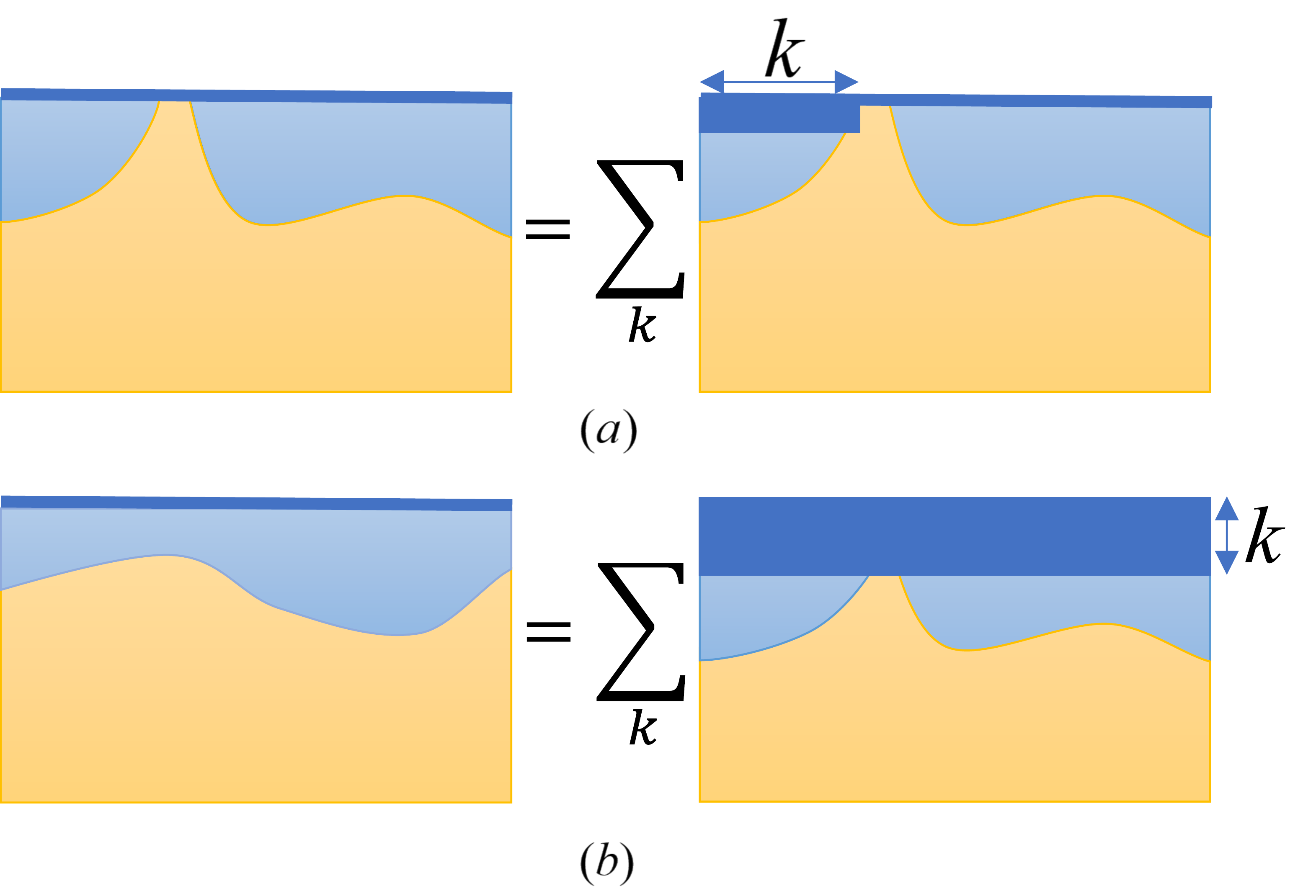}
    \caption{Relations used to calculated $\mathcal{Z}_{bb}$. Blue spins are $(12)$ and yellow spins are $id=(1)(2)$. }
    \label{fig-Zbb}
\end{figure}
To calculate $\mathcal{Z}_{bb}$, denote $w$ to be the summation of configurations where domain walls are attached to the upper boundary for at least one segment, see \reffig{fig-Zbb}(a). By classifying the position of the first attachment point, we have:
\begin{equation}\label{eq-iteZbb}
w=\sum_{k=0}^{L/2}(1-p)^{2k}\frac{1}{d}v'_k\frac{1-(1-p)^2}{d^2}v_{L/2-k-2}=\frac{1-(1-p)^L}{d^L}.
\end{equation}
Noticing \reffig{fig-Zbb}(b), we have:
\begin{equation}\label{eq-sumZbb}
    \mathcal{Z}_{bb}=\sum_{k=0}^{\infty}(1-p)^{kL}w=\frac{1}{d^L}.
\end{equation}
Alternatively, one can regard this result as the limit $n\to L/2$ in \refeq{eq-iteun} by taking both endpoints to the boundaries. The result is consistent with \refeq{eq-Zbblimit0}.

\subsection{Area Law}
The argument for area law works as follows: due to the at most linear growth, the mutual information $I(A\colon B)$ at time $t^*$ can at most be $O(t^*)$. After $t^*$ timescale, the system is trivialized and we anticipate that $I(A\colon B)$ is at most $O(1)$. Therefore, the maximal value for $I(A\colon B)$ can at most be $O(t^*)$. However, as shown in the main text, three terms $S(A)$, $S(B)$, $S(AB)$ appeared in $I(A\colon B)$ are all extensive:  they are $O(L)$ at any time. Therefore, $I(A\colon B)\rvert_{t>t^*}<O(1)$ requires a delicate cancellation. In the following, we argue that this cancellation is quite natural. 

For simplicity we set $|A|=|B|=L/2$. We anticipate:
\begin{align}
      S(AB)(t)-S(AB)\rvert_{t=\infty}=-Lf_1(L,t),\\
    S(A)(t)-S(A)\rvert_{t=\infty}=-\frac{L}{2}f_2(L,t)+f_3(L,t).  
\end{align}
Here, $f_i(L,t)$ are almost $L$ independent; $f_3(L,t)$ are bounded by some $O(1)$ value  if $t>O(t^*)$. The absence of extra term in the first equation is due to the translational invariance. The fact of $O(1)$ thermalization timescale means
\begin{equation}\label{eq-arguapp}
      f_i(L,t)\sim C_ie^{-t/t_i},
\end{equation}
where $C_i\in\mathbb{R}$ and $t_i\in\mathbb{R}_{>0}$ are some $O(1)$ value [$t_1=O(t^*)$]. By the small incremental result and nonnegativity of mutual information, we know rigorously:
\begin{equation}
    -O(1)<[f_1(L,t)-f_2(L,t)]L<C_3t-O(1) \text{~~for~}t>O(t^*).
\end{equation}

A feature of the function $g(t)=C_1e^{-t/t_1}-C_2e^{-t/t_2}$ is that it has at most one stationary point $t_*$ in $[0,+\infty)$,
which is by definition $L$ independent (just ignore $t_*$ if it does not exist). $g(t)$ monotonically approach 0 after $t_*$:
\begin{dmath}\label{eq-decay}
    |g(t)|\leq |g(t_*)|\leq \max\{|g(t_*)|,|g(t^*)|\} \text{~~for~}t>O(t_*),
\end{dmath}
hence for $t>O(\max\{t_*,t^*\})$:
\begin{equation}
\begin{aligned}
    |g(t)|L&\leq \max\{|g(t_*)|L,|g(t^*)|L\}\\
    &\leq\max\{|C_3t_*-O(1)|,|C_3t^*-O(1)|,|O(1)|\}\\
    =&O(1).
\end{aligned}
\end{equation}
The last equation is because $t_*$ and $t^*$ are $L$-independent. Due to \refeq{eq-arguapp} it is natural to expect that $f_2(L,t)-f_1(L,t)$ satisfies similar property as \refeq{eq-decay}, perhaps with some extra constant. Thus, $I(A\colon B)<O(1)$ for $t$ larger than some $L$-independent value.

The area law can also be understood assuming horizontally directed domain walls are dominant when $t>O(t^*)$. Generalizing \refeq{eq-iteZbb} to finite time, we have:
\begin{equation}\label{eq-iteun-t}
\begin{aligned}
    u_n(t)=&\frac{p}{d^2}u_{n-1}(t)+\frac{p}{d^4}\sum_{k=1}^{n-1}(1-p)^ku_{k-1}(t\!-\!1)u_{n-1-k}(t)\\
    &+\frac{(1-p)^n}{d^2}u_{n-1}(t-1).
\end{aligned}
\end{equation}
Here, $u_n(t)$ is the summation of ``hanging" configurations, under the restriction that the depth of the statistical mechanics system is $t$ [with $t<\infty$, the domain wall can disappear at the lower boundary and then reappear at a different point; to clarify, $u_n(t)$ is actually a decreasing function of $t$ due to the decrease of domain wall length although there are more configurations at larger $t$]. The boundary conditions are:
\begin{equation}
    u_{n\geq 0}(t=0)=d^2,~~u_0(t\geq 1)=1.
\end{equation}
Although we do not have an analytical expression for $u_n(t)$, the following statement can be numerically checked for $t>O(t^*)$:
\begin{equation}
    \frac{u_{0}(t)u_{L/2}(t)}{u_{L/4}(t)^2}<O(1),
\end{equation}
and this implies the area law.

\section{More on the Probabilistic Trace Setup}
\subsection{Stat-Mech Model}\label{sec-probtracemap}
This setup can be mapped to a statistical mechanics model similarly.

First,
\begin{equation}
\Bc=\begin{cases}
		\frac{1}{d^2}\Btr, & \text{with prob=$p$}\\
		\Buu, & \text{with prob=$1-p$ and $U$ random}
		\end{cases}.
\end{equation}
Hence we have:
\begin{dmath}
\mathbb{E}_C\left(\Qc\right)=\frac{p}{d^{2Q}}\Qtr+(1-p)\mathbb{E}_U\left(\Quu\right)
  =\frac{p}{d^{2Q}}\Qpermid+(1-p)\sum_{g_1,g_2\in S_{2n}}W_Q(g_1^{-1}g_2)\Qpermg.
\end{dmath}
Here the orange blocks are Haar random unitaries $U$ and $U^\dagger$; we use dashed lines for trace because we may or may not apply it; $W_Q()$ is the Weingarten function. 

Following similar calculations toward \refeq{eq-honey}, we obtain the spin model on the honeycomb lattice with the following weights:
\begin{equation}
    \prod_v\left[\frac{p}{d^{2Q}}\delta_{g_1}\delta_{g_2}+(1-p)W
    _Q(g_1^{-1}g_2)\right]\prod_h d^{Q-|g_1^{-1}g_2|}.
\end{equation}
See comments around \refeq{eq-honey} for notations.

Then we again integrate over the upper spins for each vertical bonds and obtain the triangle weights:
\begin{equation}\label{eq-Jpformula}
\begin{aligned}
          &\Tabc =\sum_{\tau\in S_{Q}}\Yabc\\
      =&\sum_{\tau\in S_{Q}} d^{-|\tau^{-1}b|-|\tau^{-1}c|}\left[p\delta_{\tau}\delta_{a}+(1-p)d^{2Q}W_Q(\tau^{-1}a)\right]\\
      =& pd^{-|b|-|c|}\delta_{a}+(1-p)J_0(a,b,c). 
\end{aligned}
\end{equation}
Here $J_0$ is exactly the triangle weight in the unitary-only case ($p=0$):
\begin{equation}
     J_0(a,b,c)=\sum_{\tau\in S_{Q}} d^{2Q-|\tau^{-1}b|-|\tau^{-1}c|}W(\tau^{-1}a).
\end{equation}

\subsection{Small Incremental}
In this setting, we can also prove that the mutual information $I(A\colon B)$ grows at most linearly for each trajectory. There are three types of effects:
\begin{itemize}
    \item For a trace channel, no matter acting inside $A$ (or $B$) or on the boundary, $I(A\colon B)$ cannot increase, due to the monotonicity of mutual information.
    \item For a unitary acting inside $A$ or $B$, $I(A\colon B)$ does not change.
    \item For a unitary acting on the boundary between $A$ and $B$, $I(A\colon B)$ can only increase a constant:
    \begin{equation}
    \begin{aligned}
    &I'(A\colon B)\\
    =&S'(A)+S'(B)-S'(AB)\\
    \leq& S'(A\backslash a)+S'(a)+S'(B\backslash b)+S'(b)-S'(AB)\\
    =&S(A\backslash a)+S'(a)+S(B\backslash b)+S'(b)-S(AB)\\
    \leq& S(A)+S(a)+S'(a)+S(B)+S(b)+S'(b)-S(AB)\\
    =&S(a)+S'(a)+S(b)+S'(b)+I(A\colon B).
    \end{aligned}
    \end{equation}
    Here two inequalities are due to the triangle inequality.
\end{itemize}
Since there is only one boundary unitary in each time step, $I(A\colon B)$ at most grows linearly.

\subsection{Numerics in Two Dimension}\label{subsec:2d}
Besides the numerical results for 1D systems discussed in the main text, we also performed simulations of a class of (2+1)D circuits. 

\begin{figure}
    \centering
\subfloat[]{\includegraphics[width=0.325\textwidth]{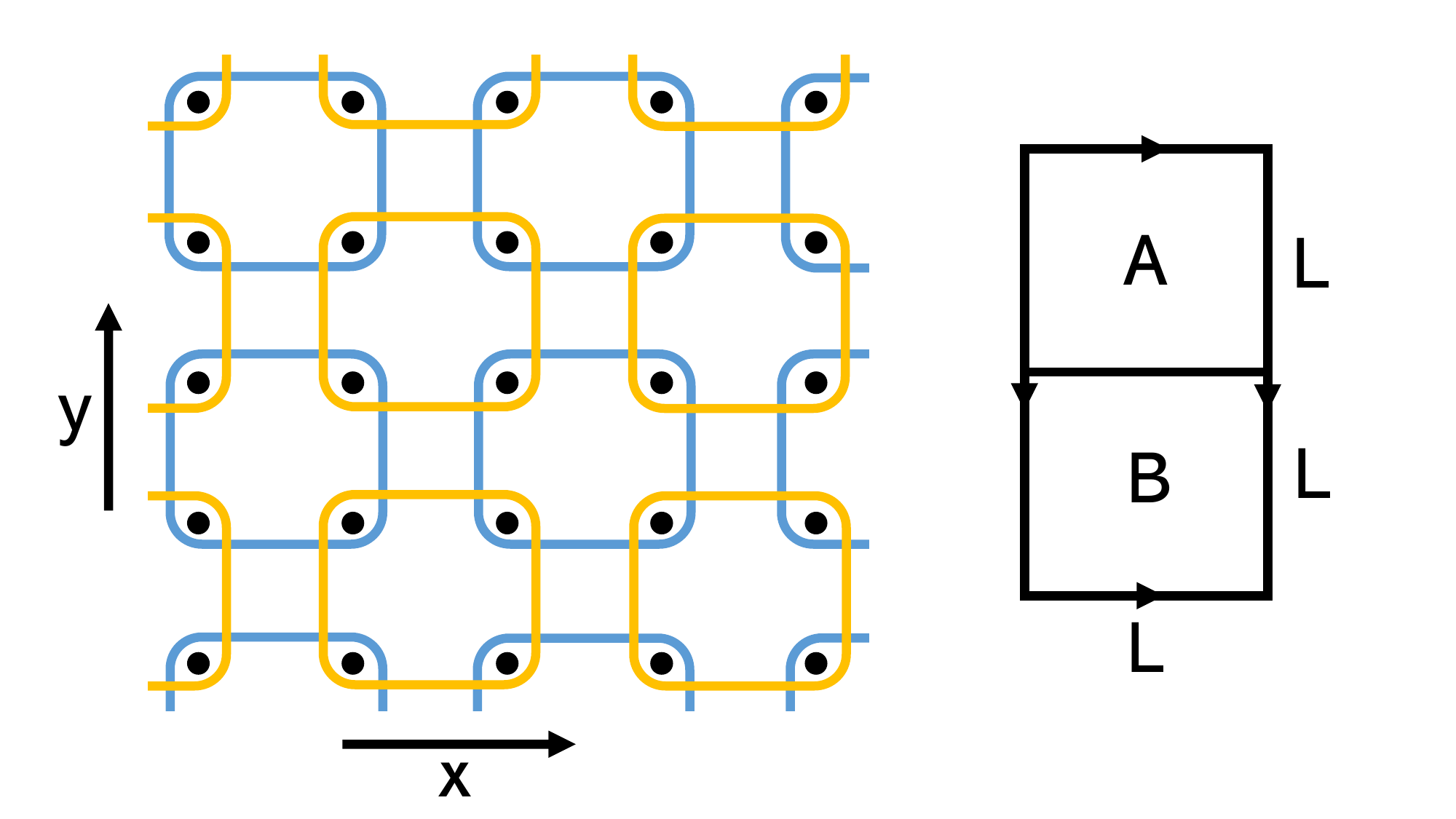}}\\
\subfloat[]{\includegraphics[width=0.325\textwidth]{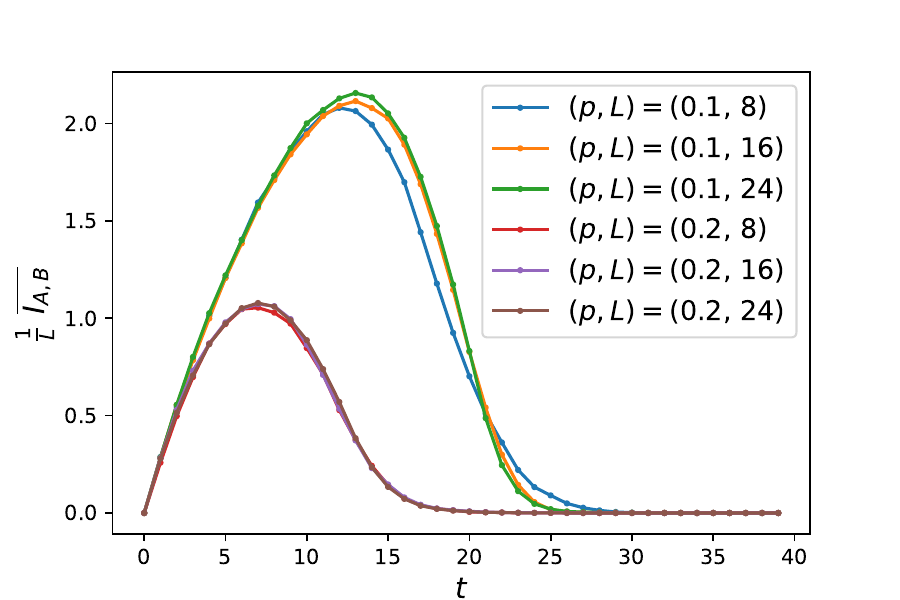}}
\caption{(a, left) An illustration of the circuit structure and (a, right) the arrangement of $A$ and $B$. See the text in Appendix \ref{subsec:2d} for a detailed description. (b) Mutual information between $A$ and $B$ for various different $p$ and $L$.}
    \label{data-2d}
\end{figure}

The circuit structure we simulated is displayed in \reffig{data-2d}(a, left): At each even (odd) time step, qubits in each blue (yellow) square are first acted by a random 4-qubit Clifford unitary gate with a probability $0.1$, then acted by a 4-qubit trace channel with a probability $0.1p$.  
The prefactors $0.1$ that appear in both probabilities are meant to ``slow down" the dynamics so that we can collect more data before the system fully thermalizes into a maximally mixed state. 
In each time step, unitary gates and measurements applied within different squares are independent of each other. The geometry of the system and the bipartition is shown in  \reffig{data-2d}(a, right): the periodic boundary condition is taken in both spatial directions, while the region $A$ and the region $B=\bar{A}$ are separated by a half-cut in the x-direction.  

Figure \ref{data-2d}(b) shows the numerical results for the mutual information between $A$ and $B$, for various different $L$ and $p$. It is clear from the plot that $I(A\colon B)$ scales linearly with the size of the boundary separating $A$ and $B$, which is proportional to $L$. Hence the area law still holds in this case (note that here the area $\propto L$).

\section{One qudit toy model}
We consider a toy model with only one qudit (with Hilbert space dimension $d$). The purpose is to illustrate what to expect in the replica calculation and large $d$ expansion.
\subsection{Random channel setup}
At each step, we apply a random quantum channel $\Phi(\cdot)=(1-p)U\cdot U^\dagger+\frac{p}{d}\mathbb{I}_1$ to the qudit. After $t$ steps, the state will be:
\begin{equation}
    (1-p)^tU\rho_0U^\dagger+\frac{1-(1-p)^t}{d}\mathbb{I}_1,
\end{equation}
where $U=U_t\cdots U_2U_1$ is again a random unitary and $\rho_0$ is a pure state.

The Rényi-$n$ entropy $\frac{1}{1-n}\log\tr(\rho^n)$ equals:
\begin{dmath}
    \frac{1}{1-n}\log\left\{\left[(1-p)^t+\frac{1-(1-p)^t}{d}\right]^n
    +(d-1)\left[\frac{1-(1-p)^t}{d}\right]^n\right\}.
\end{dmath}
The von Neumann entropy $-\tr(\rho\log\rho)$ equals:
\begin{equation}
\begin{aligned}
&-\left[(1-p)^t+\frac{1-(1-p)^t}{d}\right]\log\left[(1-p)^t+\frac{1-(1-p)^t}{d}\right]\\
&-(d-1)\frac{1-(1-p)^t}{d}\log\frac{1-(1-p)^t}{d}.
\end{aligned}
\end{equation}
Taking the large $d$ limit, the Rényi-$n$ entropy becomes: 
\begin{equation}\label{eq-toyRenyin}
   \frac{nt}{1-n}\log(1-p)+o(1),
\end{equation}
and the von Neumann entropy becomes:
\begin{dmath}
    [1-(1-p)^t]\log d-(1-p)^t\log(1-p)^t-[1-(1-p)^t]\log[1-(1-p)^t]+o(1).
\end{dmath}

We see that two limits $n\to 1$ and $d\to\infty$ do not commute. Therefore, with large $d$ expansion, one cannot calculate the von Neumann entropy by replica trick [namely, taking the limit of $n\to 1$ in \refeq{eq-toyRenyin}]. The best thing one can do is the Rényi-$n$ entropy with $n>1$.

\subsection{Probabilistic trace setup}
We apply the trace with probability $p$ at each step. Then after $t$ steps, the system remains is pure with probability $(1-p)^t$ and is maximally mixed with probability $1-(1-p)^t$.

The averaged Rényi-$n$ entropy and von Neumann entropy are equal (each trajectory is a stabilizer state):
\begin{equation}\label{eq-toy2averageS}
        \frac{1}{1-n}\overline{\log\tr(\rho^n)}=[1-(1-p)^t]\log d,
\end{equation}
while the logarithmic of averaged partition function equals:
\begin{equation}
   \frac{1}{1-n}\log\overline{\tr(\rho^n)}= \frac{1}{1-n}\log\left[ (1-p)^t+\frac{1-(1-p)^t}{d^{n-1}} \right].
\end{equation}
Taking the large $d$ limit, it becomes:
\begin{equation}
    \frac{t}{1-n}\log(1-p)+o(1).
\end{equation}
We see that the averaging over trajectories and the logarithmic do not commute.

\end{document}